\documentclass[floatfix,amsmath,amssymb,superscriptaddress,nofootinbib,longbibliography,twocolumn]{revtex4-2}

\usepackage{lipsum}
\usepackage{verbatim}
\usepackage[colorlinks=true,linkcolor=blue,citecolor=blue]{hyperref}
\usepackage[utf8]{inputenc}
\usepackage[T1]{fontenc}

\usepackage{array}
\usepackage{mathtools}

\usepackage{physics}
\usepackage{algorithmic}
\usepackage{algorithm}
\usepackage{siunitx}

\usepackage{graphicx}
\usepackage{tikz}

\usepackage{tabularx}
\usepackage{booktabs}
\newcolumntype{L}[1]{>{\raggedright\let\newline\\\arraybackslash\hspace{0pt}}m{#1}}
\newcolumntype{C}[1]{>{\centering\let\newline\\\arraybackslash\hspace{0pt}}m{#1}}
\newcolumntype{R}[1]{>{\raggedleft\let\newline\\\arraybackslash\hspace{0pt}}m{#1}}

\usepackage[caption=false]{subfig}
\newcommand{\phantomsubfloat}[1]{
    {%
        \captionsetup[subfigure]{labelformat=empty}
        \subfloat[][]{#1}
    }%
}

\usepackage{cleveref}
\crefname{equation}{Eq.}{Eqs.}
\crefname{section}{Sec.}{Secs.}
\crefname{subsection}{Sec.}{Secs.}
\crefname{appendix}{Appendix}{Appendices}
\crefname{figure}{Figure}{Figures}
\crefname{table}{Table}{Tables}
\newcommand{\methods}{\hyperref[sec:methods]{Methods}}
\newcommand{\inertlink}{}

\DeclarePairedDelimiter\ceil{\lceil}{\rceil}
\DeclarePairedDelimiter\floor{\lfloor}{\rfloor}

\DeclareMathOperator*{\argmin}{argmin}

\let\var\relax
\DeclareMathOperator*{\var}{var}

\usepackage[final]{pdfpages}

\makeatletter
\AtBeginDocument{\let\LS@rot\@undefined}
\makeatother

\begin{document}

\preprint{}
\title{Experimental Realization of a Measurement-Induced Entanglement Phase Transition on a Superconducting Quantum Processor}

\author{Jin Ming Koh}
\affiliation{Division of Physics, Mathematics and Astronomy, California Institute of Technology, Pasadena, California 91125, USA}

\author{Shi-Ning Sun}
\affiliation{Division of Engineering and Applied Science, California Institute of Technology, Pasadena, CA 91125, USA}

\author{Mario Motta}
\affiliation{IBM Quantum, IBM Research Almaden, San Jose, CA 95120, USA}

\author{Austin J.~Minnich}
\email{aminnich@caltech.edu}
\affiliation{Division of Engineering and Applied Science, California Institute of Technology, Pasadena, CA 91125, USA}

\maketitle
\date{\today}

{\bf{Ergodic quantum many-body systems undergoing unitary dynamics evolve towards increasingly entangled states characterized by an extensive scaling of entanglement entropy with system volume \cite{cabrese2005evolution, kim2013ballistic, liu2014entanglement, kaufman2016quantum, nahum2017quantum, keyserlingk2018operator}. At the other extreme, quantum systems repeatedly measured may be stabilized in a measurement eigenstate, a phenomenon known as the quantum Zeno effect \cite{davies1976quantum, misra1977zeno, wheeler1983quantum, zhu2011quantum}. Recently, the intermediate regime in which unitary evolution is interspersed with quantum measurements has become of interest \cite{elliott2015multipartite, dhar2016measurement, mazzucchi2016quantum, li2018quantum}. Numerical studies have reported the existence of distinct phases characterized by volume- and area-law entanglement entropy scaling for infrequent and frequent measurement rates, respectively, separated by a critical measurement rate \cite{chan2019unitary, li2019measurement, skinner2019measurement, szyniszewski2019entanglement, zabalo2020critical, nahum2021measurement}. The experimental investigation of these dynamic quantum phases of matter on near-term quantum hardware is challenging due to the need for repeated high-fidelity mid-circuit measurements and fine control over the evolving unitaries. Here, we report the realization of a measurement-induced entanglement transition on superconducting quantum processors with mid-circuit readout capability. We directly observe extensive and sub-extensive scaling of entanglement entropy in the volume- and area-law phases, respectively, by varying the rate of projective measurements. We further demonstrate phenomenological critical behavior of the transition by performing a data collapse for different system sizes. Our work paves the way for the use of mid-circuit measurement as an effective resource for quantum simulation on near-term quantum computers, for instance by facilitating the study of dynamic and long-range entangled quantum phases \cite{tantivasadakarn2021long, sang2021measurement}}.}

\begin{figure*}[!t]
    \centering
    \includegraphics[width=0.95\linewidth]{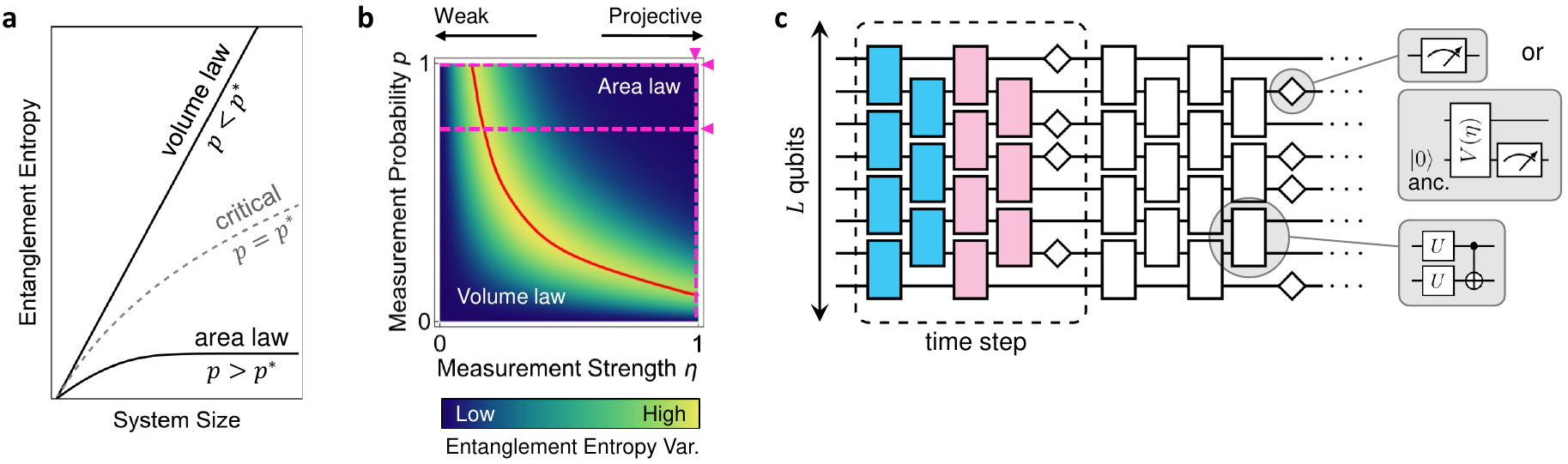}
    \phantomsubfloat{\label{fig:schematics-line}}
    \phantomsubfloat{\label{fig:schematics-space}}
    \phantomsubfloat{\label{fig:schematics-circuit}}
    \vspace{-0.5\baselineskip}
    \caption{\textbf{Measurement-induced entanglement transition using a hybrid random quantum circuit model.} 
    \textbf{(a)} Schematic of entanglement entropy versus system size, illustrating the various measurement-induced phases. Under projective measurements ($\eta = 1$), an entanglement transition occurs at a critical measurement rate $p^*$, separating a volume-law phase for $p < p^*$ and an area-law phase for $p > p^*$. In the two phases, entanglement entropy scales extensively and sub-extensively, respectively. \textbf{(b)} Qualitative diagram of steady-state entanglement entropy variance in $p$-$\eta$ space, exhibiting a ridge separating volume- and area-law regimes. Dashed magenta horizontal and vertical lines mark the parameter sweeps probed in this study. \textbf{(c)} Structure of random quantum circuits employed in this work, comprising several time steps. Each time step consists of two layers (dark blue and light pink) of randomized $2$-qubit gates in a brickwork pattern and a layer of measurements randomly placed on each qubit with probability $p$. The measurements can be projective, or weak with strength $\eta$, achieved through coupling with an ancillary qubit. Each $2$-qubit gate comprises random single-qubit rotations and a randomly-directed CX.}
    \label{fig:schematics}
\end{figure*}

The phenomenon of measurement-induced entanglement transitions stems from a competition between unitary evolution and quantum measurements, which respectively generate and destroy entanglement \cite{chan2019unitary, li2019measurement, skinner2019measurement}. The phases are illustrated schematically in \cref{fig:schematics-line}. At low measurement rates ($p$) or at small measurement strengths ($\eta$), beginning from a separable state, ballistic growth in the entanglement entropy of a subsystem first occurs followed by saturation at an average value that scales extensively with the system size. A sufficiently large measurement rate $p$ or strength $\eta$, however, can suppress entanglement. In such regimes, no ballistic growth occurs, and entanglement entropy at saturation scales sub-extensively. These dynamical features characterize the volume- and area-law phases, respectively.

Under projective measurements ($\eta = 1$), in the thermodynamic limit of infinite system size, a sharp entanglement transition occurs at a critical measurement rate $p^*$, separating the area-law and volume-law phases. Near the transition, the entanglement entropy scales logarithmically with system size to leading order. The volume- and area-law phases manifest at $p < p^*$ and $p > p^*$, respectively. In $p$--$\eta$ parameter space, the crossover between the two phases has been proposed to be identified by a ridge of increased entanglement entropy variance, as the system fluctuates between entanglement growth and suppression (see Figure~3 of Ref.~\cite{szyniszewski2019entanglement}), schematically illustrated in \cref{fig:schematics-space}. Other signatures of the transition have also been considered using measures of entanglement such as bipartite and tripartite mutual information \cite{skinner2019measurement, li2019measurement, zabalo2020critical, lunt2021measurement}. To date, a variety of models exhibiting entanglement transitions have been studied numerically, including random quantum circuits with local Clifford and Haar unitaries \cite{zabalo2020critical, szyniszewski2019entanglement, li2019measurement, skinner2019measurement} and higher-dimensional connectivity \cite{skinner2019measurement, turkeshi2020measurement, nahum2021measurement, yu2022measurement}, interacting bosonic and spin chains \cite{tang2020measurement, turkeshi2021measurement}, and symmetric circuits supporting topological phases \cite{lavasani2021measurement}. Theoretical connections to quantum error correction \cite{choi2020quantum, li2021statistical, fan2021self} and conformal field theory \cite{vasseur2019entangling, bao2020theory, jian2020measurement, sang2021entanglement, block2022measurement}, concerning the stability of the phases and criticality, have also been uncovered.

Experimentally realizing these dynamic quantum phases of matter and the transition between them on near-term quantum hardware, however, is difficult due to the need for repeated high-fidelity measurements ($\gtrsim 10$) over non-trivial system sizes ($\gtrsim 10$ qubits). Although the evolving unitaries are random, fine control over their distribution and entangling properties is necessary for quantitative agreement with theoretical expectations, as gate infidelity can distort the transition boundary and presents a confounding source of entropy. In a recent work on a trapped-ion quantum computer, a purification phase transition, expected to occur concurrently with entanglement phase transitions \cite{gullans2020dynamical, bao2020theory}, was observed using Clifford circuits and deferred measurements \cite{noel2021observation}. However, the ancillary qubit overhead of this approach complicated the investigation of larger system sizes.


We leveraged recent hardware advances in superconducting quantum computers, particularly the ability to perform sub-microsecond ($\sim \SI{750}{\nano\second}$) mid-circuit measurements \cite{corcoles2021exploiting}, to probe a volume- to area-law entanglement phase transition on a non-Clifford random circuit model. Our experiments spanned $\sim 5200$ hardware device-hours ($\sim 32000$ qubit-hours) over multiple quantum processors, making them among the most resource-intensive quantum simulations on near-term hardware reported to date. The hybrid random circuit model used to realize measurement-induced entanglement transitions is shown in \cref{fig:schematics-circuit}. An experiment circuit comprises interleaved unitary entangling layers and mid-circuit measurements on an open 1D qubit chain, repeated over several time steps. Unlike in prior experimental works probing measurement-induced phase transitions, measurements are not deferred. The unitary layers contain randomized $2$-qubit gates in a brickwork pattern, each comprising Haar-uniform single-qubit rotations and a randomly-directed CX gate. In the measurement layer, measurements are placed independently on each qubit with probability $p$. These measurements are projective ($\sigma^z$) or null-type weak \cite{gebhart2020topological, zilberberg2013null}, the latter implemented by coupling the system qubit to an ancillary qubit through a unitary $V(\eta)$ before measuring the ancilla. Weak-limit ($\eta = 0$) measurements do not affect the system, and $\eta = 1$ coincides with the projective limit, with intermediate $\eta$ smoothly interpolating between the extremes. The particular structure of our circuits was designed to minimize circuit depth (see \methods). 

After the evolution is complete, we characterize the entanglement of the state by measuring R{\'e}nyi entanglement entropies
\begin{equation}\begin{split}
    S_\alpha = \frac{1}{1 - \alpha} \log_2 \left[\Tr \left(\rho_A^\alpha\right)\right],
    \label{eq:renyi-entanglement-entropy}
\end{split}\end{equation}
at order $\alpha \geq 0$, with the reduced density matrix $\rho_A$ of a subsystem $A$ recovered through quantum state tomography (QST) for each recorded trajectory defined by the mid-circuit measurement outcomes. Owing to the substantial cost of computing entanglement measures such as bipartite and tripartite mutual information, we characterize the transition through mean entropy $\expval{S_\alpha}$ and variance $\var(S_\alpha)$ computed over random samples of the quantum circuits. The sampling and post-selection of distinct trajectories is resource-intensive, with an underlying exponential scaling in the number of mid-circuit measurements. To reduce the computational cost and render the experiments feasible, we perform simultaneous tomography measurements on sets of mutually unbiased bases (MUBs), which are groups of commuting Pauli strings of maximal size (see \methods). 

\begin{figure*}[!t]
    \centering
    \includegraphics[width=\linewidth]{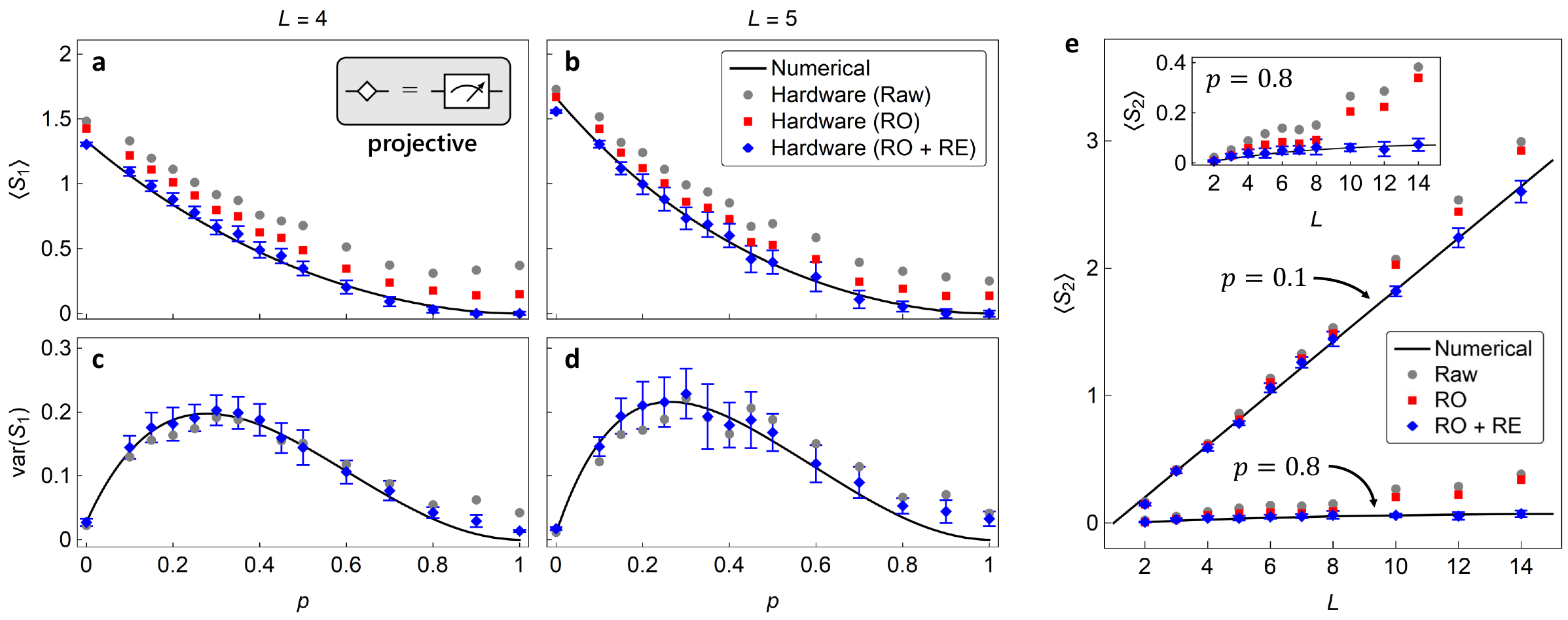}
    \phantomsubfloat{\label{fig:results-proj-trans-L4-mean}}
    \phantomsubfloat{\label{fig:results-proj-trans-L5-mean}}
    \phantomsubfloat{\label{fig:results-proj-trans-L4-var}}
    \phantomsubfloat{\label{fig:results-proj-trans-L5-var}}
    \phantomsubfloat{\label{fig:results-proj-scaling}}
    \vspace{-2\baselineskip}
    \caption{\textbf{Entanglement crossover and system size scaling under projective measurements.} Average von Neumann entanglement entropy $\expval{S_1}$ versus measurement rate $p$ at system sizes \textbf{(a)} $L = 4$ and \textbf{(b)} $L = 5$, and $S_1$ entanglement entropy variance versus $p$ at \textbf{(c)} $L = 4$ and \textbf{(d)} $L = 5$, obtained on quantum hardware with readout error mitigation (RO) and residual entropy correction (RE), shown as blue diamonds. Raw hardware data (gray dots), data with RO applied (red squares), and ideal noiseless numerical results without sampling (solid line) are shown for comparison. Excellent agreement between hardware data with RO and RE applied and ideal entropies is observed, and the characteristic ridge of high entropy variance indicative of a crossover is clearly visible. \textbf{(e)} Second-order R{\'e}nyi entanglement entropy $\expval{S_2}$ versus $L$, up to $L = 14$ qubits, obtained on hardware. At $p = 0.1 < p^*$, the system exhibits volume-law behavior and entanglement entropy is linearly proportional to system size; whereas at high $p = 0.8 > p^*$ the system exhibits area-law scaling  and the entanglement entropy saturates. Inset: zoomed-in plot of $\expval{S_2}$ versus $L$ for $p = 0.8$ data. Quantum devices \{\textit{ibm\_lagos}, \textit{ibm\_perth}, \textit{ibmq\_jakarta}, \textit{ibmq\_casablanca}\} were used for $L \leq 5$, and \{\textit{ibm\_hanoi}, \textit{ibm\_cairo}, \textit{ibm\_kolkata}, \textit{ibm\_auckland}, \textit{ibm\_washington}\} with sub-microsecond readout were used for $L > 5$. Error bars reflect $90\%$ confidence intervals estimated from statistical bootstrapping.}
    \label{fig:results-proj}
\end{figure*}

The mean and variance of steady-state von Neumann entanglement entropy $S_1$ versus measurement rate $p$ on quantum hardware at system sizes $L = 4, 5$ and subsystem $\abs{A} = \floor{L / 2}$ under mid-circuit projective measurements are shown in \cref{fig:results-proj-trans-L4-mean,fig:results-proj-trans-L5-mean,fig:results-proj-trans-L4-var,fig:results-proj-trans-L5-var}. In \cref{fig:results-proj-trans-L4-mean,fig:results-proj-trans-L5-mean}, we observe a monotonic decrease in $\expval{S_1}$ with increasing $p$, consistent with the suppression of entanglement with increasing measurement rate. Additionally, from \cref{fig:results-proj-trans-L4-var,fig:results-proj-trans-L5-var}, a peak is evident in $\var{(S_1)}$ near $p \approx 0.25$, suggestive of a crossover between a volume-law (area-law) phase for $p < p^*$ ($p > p^*$). 

We note that the raw experiment data consistently overestimates entanglement entropy due to entropic contributions from two sources: the misassignment of mid-circuit measurement outcomes due to readout error, and an effective decoherence-like effect, arising from gate errors averaged over the random circuits. To mitigate these non-idealities, two error mitigation schemes are employed. First, readout error mitigation (RO) reduces the effect of measurement bit-flip errors \cite{jurcevic2021demonstration, kandala2017hardware}. Second, a residual entropy correction (RE) is applied by noting that the entanglement entropy at $p = 1$ should vanish, but that detected on hardware is nonzero. An approximate correction is obtained by subtracting this residual entropy from all data at the same $L$ (see \methods). While the raw experiment data exhibits already a qualitative agreement with theory, RO and RE combined bring the measured mean and variance of $S_1$ into quantitative agreement. We remark that error bars and deviations from ideal noiseless numerics are more prominent at larger system size $L = 5$ than $L = 4$, a consequence of the deeper circuits and a reduced ability to avoid qubits with larger error rates.

To directly demonstrate the realization of distinct volume- and area-law entanglement phases, we probe the scaling of entanglement entropy with system size. Taking $p = 0.1 < p^*$ below the crossover and $p = 0.8 > p^*$ above, we obtain hardware data on up to $L = 14$ qubits using a quarter subsystem size (see \methods). These simulations were performed on $27$- and $127$-qubit quantum processors with sub-microsecond ($\approx \SI{750}{\nano\second}$) readout and $T_1, T_2 \approx \SI{100}{\micro\second}$ decoherence times. The MUB-based QST, which is crucial in managing the resources needed for these experiments (see \methods), in principle allows the recovery of entanglement entropy at any order $\alpha$, but higher $\alpha$ is more susceptible to hardware noise. Present hardware capabilities are nonetheless sufficiently advanced to afford probing $\alpha > 1$. To demonstrate this flexibility, we report steady-state second-order R{\'e}nyi entanglement entropy $\expval{S_2}$ versus $L$ in \cref{fig:results-proj-scaling}. At $p = 0.1$, linear scaling of $\expval{S_2}$ with system size $L$ is evident; in contrast, $\expval{S_2}$ scales sub-linearly at $p = 0.8$, rapidly saturating and becoming largely independent of system size. These scaling characteristics are precisely those expected in the volume- and area-law phases. We further remark that such scaling behavior extends to all higher orders of entanglement entropy---as every $S_{\alpha > 1}$ differs from $S_\infty$ by at most a constant factor \cite{skinner2019measurement}, they must all exhibit similar scaling with system size. These results thus provide direct evidence of a volume- to area-law quantum phase transition realized on quantum hardware. 

\begin{figure*}[!t]
    \centering
    \includegraphics[width=0.58\linewidth]{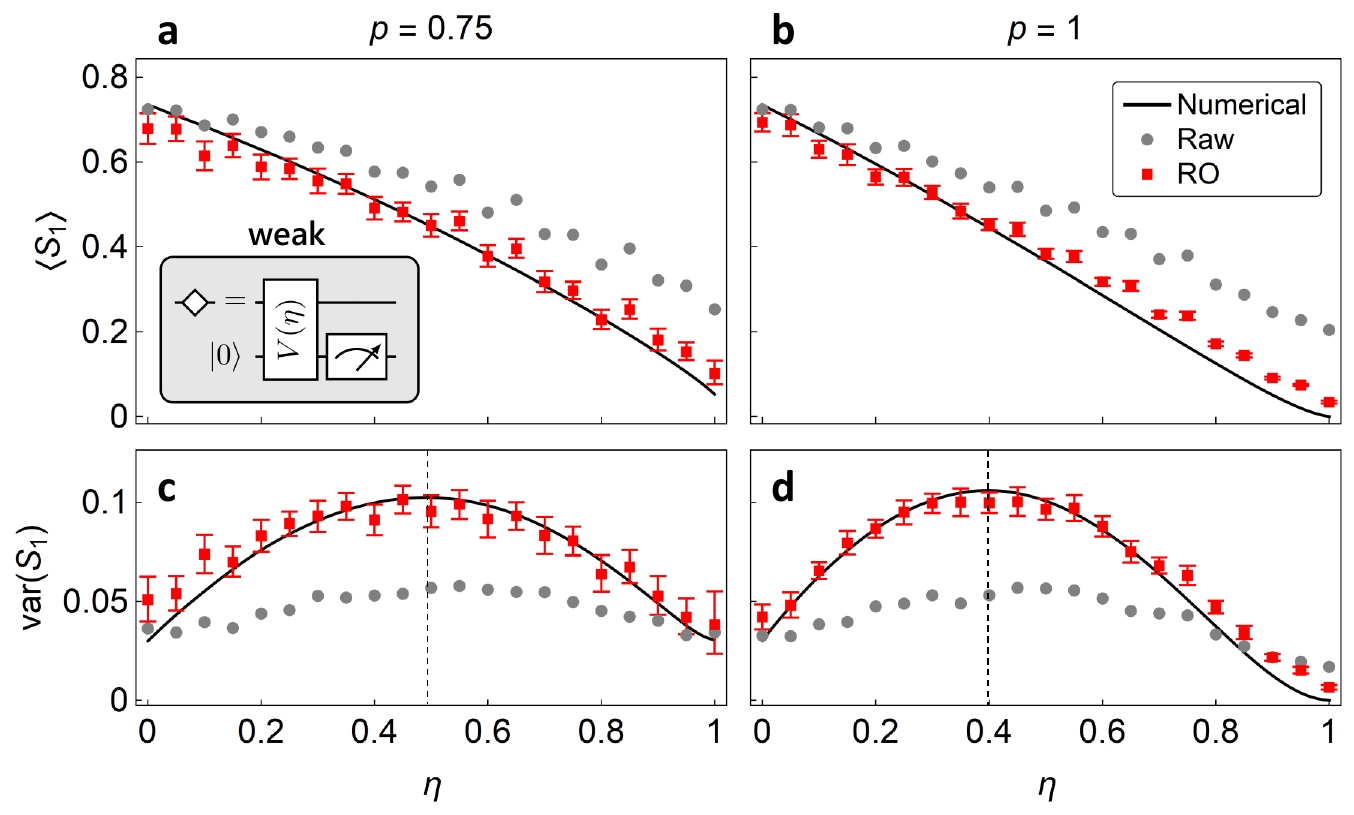}
    \phantomsubfloat{\label{fig:results-weak-p0750-mean}}
    \phantomsubfloat{\label{fig:results-weak-p1000-mean}}
    \phantomsubfloat{\label{fig:results-weak-p0750-var}}
    \phantomsubfloat{\label{fig:results-weak-p1000-var}}
    \vspace{-2\baselineskip}
    \caption{\textbf{Entanglement crossover under weak measurements.} \textbf{(a)-(b)} Average von Neumann entanglement entropy and \textbf{(c)-(d)} variance at $p = 0.75$ and $p = 1$ versus measurement strength, obtained on quantum hardware with readout error mitigation (RO) applied, shown as red squares. Raw hardware data (gray dots) and ideal noiseless numerical results without sampling (solid line) are shown for comparison. A ridge of high entropy variance indicative of a crossover is visible, similar to the projective case; the location of the crossover (vertical dashed line) is now dependent on $p$. Quantum devices \{\textit{ibmq\_guadalupe}, \textit{ibmq\_montreal}\} were used. Error bars reflect $98\%$ confidence intervals estimated from statistical bootstrapping.}
    \label{fig:figure-results-weak}
\end{figure*}

Beyond projective measurements, we demonstrate an entanglement crossover using null-type weak measurements. Though weak measurements do not cause complete quantum state collapses, they nonetheless provide partial classical information on the measured system and can reduce entanglement entropy, likewise serving competitively against the evolving unitaries that generate entanglement. We report $\expval{S_1}$ versus measurement strength $\eta$ in \cref{fig:results-weak-p0750-mean,fig:results-weak-p1000-mean} for $p = 0.75$ and $p = 1$, respectively, at system size $L = 3$. These parameter sweeps represent horizontal slices of the $p$--$\eta$ diagram of \cref{fig:schematics-space}, complementary to vertical slicing in the projective case. Only RO mitigation was applied as cumulative hardware errors were sufficiently small to make RE corrections negligible. At fixed $p$, $\expval{S_1}$ decreases monotonically with $\eta$, as expected from the entanglement suppression of increasingly strong measurements. As shown in \cref{fig:results-weak-p0750-var,fig:results-weak-p1000-var}, the signature peak in entanglement entropy variance is likewise evident, marking a crossover between volume- to area-law phases. This crossover occurs at larger $\eta$ when $p = 0.75$ than when $p = 1$, reflecting a trade-off between measurement rate $p$ and strength $\eta$ necessary to cause a transition. Such a trade-off is also clear in \cref{fig:schematics-space} as the measurements depart from the projective limit.

In both projective (\cref{fig:results-proj-trans-L4-var,fig:results-proj-trans-L5-var}) and weak measurement (\cref{fig:results-weak-p0750-var,fig:results-weak-p1000-var}) experiments, we note slight deviations in entanglement entropy variance at large $p \gtrsim 0.9$ and $\eta \gtrsim 0.9$, where $\var{(S_1)}$ diminishes close to zero, but the hardware data does not decrease as steeply. We attribute this discrepancy to variations in hardware noise uncorrected by RO and RE such as fluctuations in readout and coherent errors over the duration of the experiment, which manifest as an additional spread in measured entanglement entropy. 

\begin{figure*}[!t]
    \centering
    \includegraphics[width=0.85\linewidth]{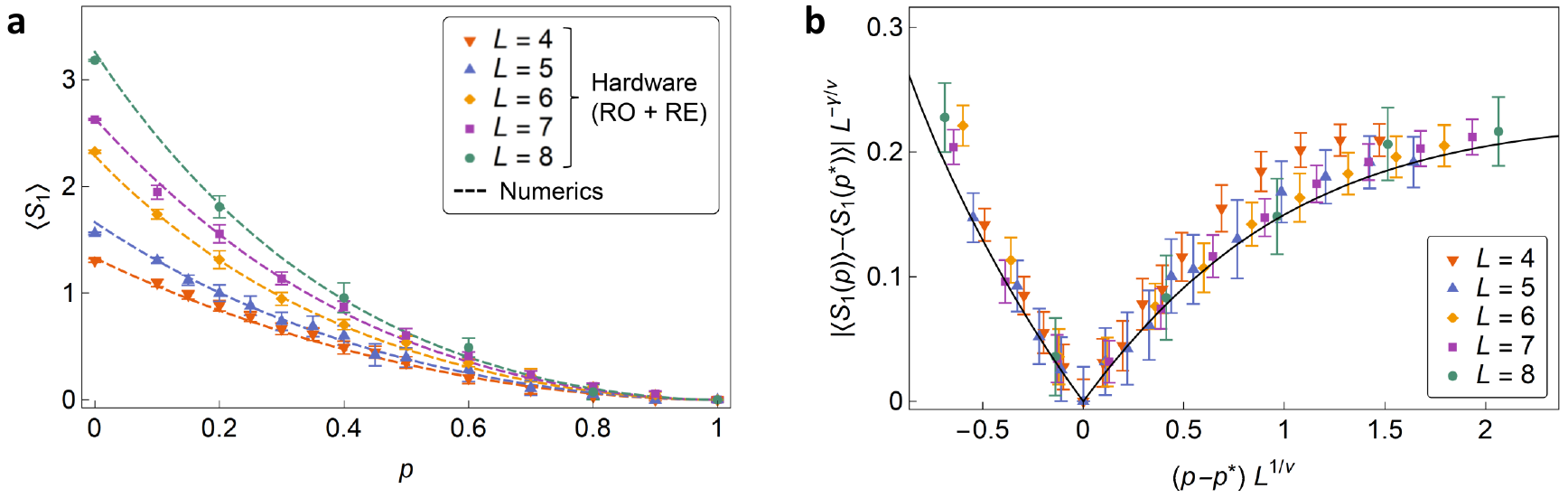}
    \phantomsubfloat{\label{fig:results-collapse-before}}
    \phantomsubfloat{\label{fig:results-collapse-after}}
    \vspace{-0.5\baselineskip}
    \caption{\textbf{Phenomenological critical behavior of the entanglement transition.} \textbf{(a)} Average von Neumann entanglement entropies versus $p$ under projective measurements for $4 \leq L \leq 8$ at half subsystem size, obtained on quantum hardware with RO and RE mitigation applied. Dashed lines are ideal noiseless numerical results for comparison. \textbf{(b)} Von Neumann entanglement entropies rescaled by a finite-size scaling form, showing a collapse of all data onto a single curve. A critical measurement rate $p^* = 0.25$ is used corresponding to the value at which the variance of $S_1$ exhibits a maximum, indicative of a crossover (\cref{fig:results-proj-trans-L5-var}). Estimated best-fit critical exponents are $\gamma \approx 1.9 \pm 0.4$ and $\nu \approx 2.1 \pm 0.3$. Solid line shows noiseless numerical results at $L = 16$, with $p$ and $\expval{S_1}$ rescaled using best-fit $(\gamma, \nu)$ from experiment, for comparison. Quantum devices \{\textit{ibm\_lagos}, \textit{ibm\_perth}, \textit{ibmq\_jakarta}, \textit{ibmq\_casablanca}\} were used for $L \leq 5$, and \{\textit{ibm\_hanoi}, \textit{ibm\_cairo}, \textit{ibmq\_kolkata}\} were used for $L > 5$. Error bars reflect $90\%$ confidence intervals estimated from statistical bootstrapping.}
    \label{fig:results-collapse}
\end{figure*}



Well-established in statistical mechanics, critical transitions are characterized by their scaling exponents, which also define universality classes. Here, we show that our observed crossovers exhibit the phenomenology of critical behavior by performing a data collapse onto a generic power-law scaling form. The specific form of the ansatz remains a topic of discussion and depends on the structure of the random circuits \cite{li2018quantum, li2019measurement, szyniszewski2019entanglement, li2021conformal}. Assuming correlation length $\xi \sim \abs{p - p^*}^{-\nu}$ and entanglement entropy $\expval{S_\alpha} \sim \abs{p - p^*}^\gamma$, we take the finite-size scaling ansatz \cite{fisher1972scaling, binder2010monte} to be
\begin{equation}\begin{split}
    \abs{\expval{S_\alpha(p)} - \expval{S_\alpha(p^*)}} L^{-\gamma / \nu} 
    &= F'\left[ (L/\xi)^{1/\nu} \right] \\
    &= F\left[L^{1/\nu} (p - p^*)\right],
\end{split}\end{equation}
for unknown scaling functions $F$, $F'$ and critical exponents $\gamma, \nu$. Thus, upon appropriate rescaling by $\gamma, \nu$, the measured entanglement entropy $\abs{\expval{S_\alpha(p)} - \expval{S_\alpha(p^*)}}$ is expected to fall on the same curve for every $L$ if the system exhibits critical behavior. We report experimentally-measured $\expval{S_1}$ against $p$ for $4 \leq L \leq 8$ in \cref{fig:results-collapse-before}. To perform the data collapse, we took $p^* \approx 0.25$, the location of the $S_1$ variance peak previously shown (\cref{fig:results-proj-trans-L5-var}), for simplicity. We then obtained best-fit estimates of $\gamma$ and $\nu$ (see \methods). As shown in \cref{fig:results-collapse-after}, upon rescaling, the experimental data collapses onto a single curve for values of $\gamma \approx 1.9 \pm 0.4$ and $\nu \approx 2.1 \pm 0.3$. A similar result is obtained if an extrapolation scheme to the thermodynamic limit is used to estimate $p^* \approx 0.22$, with nearly identical values of $\gamma, \nu$ exponents. A qualitatively similar collapse occurs even with a weakened version of RE correction (see Supplementary Note~\inertlink{5}).

We emphasize that the collapse procedure is dependent only on hardware-derived experiment data and operates entirely within the context of an unknown scaling function $F$, with no external reference. We note that as system sizes $L \leq 8$ were limited by feasibility on hardware, finite-size effects, which manifest as $L$-dependent distortions from the scaling form, may play a role in our experiments. We verify that the collapsed experiment data is consistent with numerics at a larger $L = 16$ system size (solid line in \cref{fig:results-collapse-after}). Further comparison of the critical behavior of the transition against previous numerical studies is not straightforward owing to the different hybrid random circuits used here and the relatively small range of $L$. Reported values of $p^*$ and critical exponents vary depending on circuit structure and scaling ansatz \cite{skinner2019measurement, szyniszewski2019entanglement, li2018quantum, li2019measurement}, and estimates from collapses of mutual information \cite{zabalo2020critical, lunt2021measurement} differ from those of entanglement entropy. Nonetheless, the tight collapse of the hardware data at different $L$ onto a single curve indicates that the phenomenology of critical behavior has been observed, highlighting the self-consistency and fidelity of our quantum circuit executions.

Our results illustrate that mid-circuit measurements on near-term quantum hardware can be effectively utilized in exploring dynamic quantum phases of matter. Future experimental studies of entanglement phase transitions may examine methods of characterizing the transition with reduced resource costs \cite{bao2020theory, gullans2020scalable}. Moreover, extensions of the paradigmatic random hybrid circuit model presently studied may harbor more intricate behavior, such as an interplay of topological and entanglement phase transitions \cite{lavasani2021measurement, sang2021measurement}, that become accessible with $\sim 10^2$ mid-circuit measurements and $\sim 0.1\%$ two-qubit gate infidelities on hardware. More generally, mid-circuit measurements can reduce the resources required to prepare novel entangled quantum states of matter \cite{lavasani2021measurement, sang2021measurement}. Our work thus paves the way for the use of hybrid quantum circuits as an effective resource to advance quantum simulation on near-term quantum hardware.

\newpage

\section*{Data Availability}

The data that support the findings of this study are available from the corresponding author upon reasonable request.

\section*{Code Availability}

The code used in this study is available from the corresponding author upon reasonable request.

\section*{Acknowledgements}

S.S. and A.J.M. were supported by the U.S. Department of Energy under Award No. DE-SC0019374. M. M. acknowledges Jody Burks, Douglas McClure, Sarah Sheldon, and Matthew Stypulkoski for help with access to, and use of, IBM Quantum devices. The authors acknowledge the use of IBM Quantum services for this work.

\section*{Competing Interests}

The authors declare no competing interests.

\newpage

\section*{Author Contributions}

J.M.K., S.-N.S. and A.J.M. conceived and initiated the project. A.J.M. supervised the project. J.M.K. developed the quantum simulation codebase, ran experiments on emulators and quantum hardware, and wrote most of the manuscript. M.M. contributed to the codebase and ran experiments on quantum hardware. J.M.K., M.M. and A.J.M. analyzed computational and experiment results. S.-N.S., M.M. and A.J.M. edited the manuscript. The manuscript reflects the contributions of all authors.

\bibliography{ref-ph, ref-qc}

\begin{thebibliography}{58}%
\makeatletter
\providecommand \@ifxundefined [1]{%
 \@ifx{#1\undefined}
}%
\providecommand \@ifnum [1]{%
 \ifnum #1\expandafter \@firstoftwo
 \else \expandafter \@secondoftwo
 \fi
}%
\providecommand \@ifx [1]{%
 \ifx #1\expandafter \@firstoftwo
 \else \expandafter \@secondoftwo
 \fi
}%
\providecommand \natexlab [1]{#1}%
\providecommand \enquote  [1]{``#1''}%
\providecommand \bibnamefont  [1]{#1}%
\providecommand \bibfnamefont [1]{#1}%
\providecommand \citenamefont [1]{#1}%
\providecommand \href@noop [0]{\@secondoftwo}%
\providecommand \href [0]{\begingroup \@sanitize@url \@href}%
\providecommand \@href[1]{\@@startlink{#1}\@@href}%
\providecommand \@@href[1]{\endgroup#1\@@endlink}%
\providecommand \@sanitize@url [0]{\catcode `\\12\catcode `\$12\catcode
  `\&12\catcode `\#12\catcode `\^12\catcode `\_12\catcode `\%12\relax}%
\providecommand \@@startlink[1]{}%
\providecommand \@@endlink[0]{}%
\providecommand \url  [0]{\begingroup\@sanitize@url \@url }%
\providecommand \@url [1]{\endgroup\@href {#1}{\urlprefix }}%
\providecommand \urlprefix  [0]{URL }%
\providecommand \Eprint [0]{\href }%
\providecommand \doibase [0]{https://doi.org/}%
\providecommand \selectlanguage [0]{\@gobble}%
\providecommand \bibinfo  [0]{\@secondoftwo}%
\providecommand \bibfield  [0]{\@secondoftwo}%
\providecommand \translation [1]{[#1]}%
\providecommand \BibitemOpen [0]{}%
\providecommand \bibitemStop [0]{}%
\providecommand \bibitemNoStop [0]{.\EOS\space}%
\providecommand \EOS [0]{\spacefactor3000\relax}%
\providecommand \BibitemShut  [1]{\csname bibitem#1\endcsname}%
\let\auto@bib@innerbib\@empty
\bibitem [{\citenamefont {Calabrese}\ and\ \citenamefont
  {Cardy}(2005)}]{cabrese2005evolution}%
  \BibitemOpen
  \bibfield  {author} {\bibinfo {author} {\bibfnamefont {P.}~\bibnamefont
  {Calabrese}}\ and\ \bibinfo {author} {\bibfnamefont {J.}~\bibnamefont
  {Cardy}},\ }\bibfield  {title} {\bibinfo {title} {Evolution of entanglement
  entropy in one-dimensional systems},\ }\href
  {https://doi.org/10.1088/1742-5468/2005/04/p04010} {\bibfield  {journal}
  {\bibinfo  {journal} {J. Stat. Mech.: Theory Exp.}\ }\textbf {\bibinfo
  {volume} {2005}}\bibinfo  {number} { (04)},\ \bibinfo {pages}
  {P04010}}\BibitemShut {NoStop}%
\bibitem [{\citenamefont {Kim}\ and\ \citenamefont
  {Huse}(2013)}]{kim2013ballistic}%
  \BibitemOpen
\bibfield  {number} {  }\bibfield  {author} {\bibinfo {author} {\bibfnamefont
  {H.}~\bibnamefont {Kim}}\ and\ \bibinfo {author} {\bibfnamefont {D.~A.}\
  \bibnamefont {Huse}},\ }\bibfield  {title} {\bibinfo {title} {Ballistic
  spreading of entanglement in a diffusive nonintegrable system},\ }\href
  {https://doi.org/10.1103/PhysRevLett.111.127205} {\bibfield  {journal}
  {\bibinfo  {journal} {Phys. Rev. Lett.}\ }\textbf {\bibinfo {volume} {111}},\
  \bibinfo {pages} {127205} (\bibinfo {year} {2013})}\BibitemShut {NoStop}%
\bibitem [{\citenamefont {Liu}\ and\ \citenamefont
  {Suh}(2014)}]{liu2014entanglement}%
  \BibitemOpen
  \bibfield  {author} {\bibinfo {author} {\bibfnamefont {H.}~\bibnamefont
  {Liu}}\ and\ \bibinfo {author} {\bibfnamefont {S.~J.}\ \bibnamefont {Suh}},\
  }\bibfield  {title} {\bibinfo {title} {Entanglement tsunami: Universal
  scaling in holographic thermalization},\ }\href
  {https://doi.org/10.1103/PhysRevLett.112.011601} {\bibfield  {journal}
  {\bibinfo  {journal} {Phys. Rev. Lett.}\ }\textbf {\bibinfo {volume} {112}},\
  \bibinfo {pages} {011601} (\bibinfo {year} {2014})}\BibitemShut {NoStop}%
\bibitem [{\citenamefont {Kaufman}\ \emph {et~al.}(2016)\citenamefont
  {Kaufman}, \citenamefont {Tai}, \citenamefont {Lukin}, \citenamefont
  {Rispoli}, \citenamefont {Schittko}, \citenamefont {Preiss},\ and\
  \citenamefont {Greiner}}]{kaufman2016quantum}%
  \BibitemOpen
  \bibfield  {author} {\bibinfo {author} {\bibfnamefont {A.~M.}\ \bibnamefont
  {Kaufman}}, \bibinfo {author} {\bibfnamefont {M.~E.}\ \bibnamefont {Tai}},
  \bibinfo {author} {\bibfnamefont {A.}~\bibnamefont {Lukin}}, \bibinfo
  {author} {\bibfnamefont {M.}~\bibnamefont {Rispoli}}, \bibinfo {author}
  {\bibfnamefont {R.}~\bibnamefont {Schittko}}, \bibinfo {author}
  {\bibfnamefont {P.~M.}\ \bibnamefont {Preiss}},\ and\ \bibinfo {author}
  {\bibfnamefont {M.}~\bibnamefont {Greiner}},\ }\bibfield  {title} {\bibinfo
  {title} {Quantum thermalization through entanglement in an isolated many-body
  system},\ }\href {https://doi.org/10.1126/science.aaf6725} {\bibfield
  {journal} {\bibinfo  {journal} {Science}\ }\textbf {\bibinfo {volume}
  {353}},\ \bibinfo {pages} {794} (\bibinfo {year} {2016})}\BibitemShut
  {NoStop}%
\bibitem [{\citenamefont {Nahum}\ \emph {et~al.}(2017)\citenamefont {Nahum},
  \citenamefont {Ruhman}, \citenamefont {Vijay},\ and\ \citenamefont
  {Haah}}]{nahum2017quantum}%
  \BibitemOpen
  \bibfield  {author} {\bibinfo {author} {\bibfnamefont {A.}~\bibnamefont
  {Nahum}}, \bibinfo {author} {\bibfnamefont {J.}~\bibnamefont {Ruhman}},
  \bibinfo {author} {\bibfnamefont {S.}~\bibnamefont {Vijay}},\ and\ \bibinfo
  {author} {\bibfnamefont {J.}~\bibnamefont {Haah}},\ }\bibfield  {title}
  {\bibinfo {title} {Quantum entanglement growth under random unitary
  dynamics},\ }\href {https://doi.org/10.1103/PhysRevX.7.031016} {\bibfield
  {journal} {\bibinfo  {journal} {Phys. Rev. X}\ }\textbf {\bibinfo {volume}
  {7}},\ \bibinfo {pages} {031016} (\bibinfo {year} {2017})}\BibitemShut
  {NoStop}%
\bibitem [{\citenamefont {von Keyserlingk}\ \emph {et~al.}(2018)\citenamefont
  {von Keyserlingk}, \citenamefont {Rakovszky}, \citenamefont {Pollmann},\ and\
  \citenamefont {Sondhi}}]{keyserlingk2018operator}%
  \BibitemOpen
  \bibfield  {author} {\bibinfo {author} {\bibfnamefont {C.~W.}\ \bibnamefont
  {von Keyserlingk}}, \bibinfo {author} {\bibfnamefont {T.}~\bibnamefont
  {Rakovszky}}, \bibinfo {author} {\bibfnamefont {F.}~\bibnamefont
  {Pollmann}},\ and\ \bibinfo {author} {\bibfnamefont {S.~L.}\ \bibnamefont
  {Sondhi}},\ }\bibfield  {title} {\bibinfo {title} {Operator hydrodynamics,
  {OTOCs}, and entanglement growth in systems without conservation laws},\
  }\href {https://doi.org/10.1103/PhysRevX.8.021013} {\bibfield  {journal}
  {\bibinfo  {journal} {Phys. Rev. X}\ }\textbf {\bibinfo {volume} {8}},\
  \bibinfo {pages} {021013} (\bibinfo {year} {2018})}\BibitemShut {NoStop}%
\bibitem [{\citenamefont {Davies}\ and\ \citenamefont
  {Davies}(1976)}]{davies1976quantum}%
  \BibitemOpen
  \bibfield  {author} {\bibinfo {author} {\bibfnamefont {E.~B.}\ \bibnamefont
  {Davies}}\ and\ \bibinfo {author} {\bibfnamefont {E.}~\bibnamefont
  {Davies}},\ }\href@noop {} {\emph {\bibinfo {title} {Quantum theory of open
  systems}}}\ (\bibinfo  {publisher} {Academic Press},\ \bibinfo {address}
  {London},\ \bibinfo {year} {1976})\BibitemShut {NoStop}%
\bibitem [{\citenamefont {Misra}\ and\ \citenamefont
  {Sudarshan}(1977)}]{misra1977zeno}%
  \BibitemOpen
  \bibfield  {author} {\bibinfo {author} {\bibfnamefont {B.}~\bibnamefont
  {Misra}}\ and\ \bibinfo {author} {\bibfnamefont {E.~C.~G.}\ \bibnamefont
  {Sudarshan}},\ }\bibfield  {title} {\bibinfo {title} {The {Zeno}’s paradox
  in quantum theory},\ }\href {https://doi.org/10.1063/1.523304} {\bibfield
  {journal} {\bibinfo  {journal} {J. Math. Phys.}\ }\textbf {\bibinfo {volume}
  {18}},\ \bibinfo {pages} {756} (\bibinfo {year} {1977})}\BibitemShut
  {NoStop}%
\bibitem [{\citenamefont {Wheeler}\ and\ \citenamefont
  {Zurek}(1983)}]{wheeler1983quantum}%
  \BibitemOpen
  \bibfield  {author} {\bibinfo {author} {\bibfnamefont {J.}~\bibnamefont
  {Wheeler}}\ and\ \bibinfo {author} {\bibfnamefont {W.}~\bibnamefont
  {Zurek}},\ }\href@noop {} {\emph {\bibinfo {title} {Quantum Theory and
  Measurement}}}\ (\bibinfo  {publisher} {Princeton University Press},\
  \bibinfo {address} {New Jersey},\ \bibinfo {year} {1983})\BibitemShut
  {NoStop}%
\bibitem [{\citenamefont {Zhu}\ \emph {et~al.}(2011)\citenamefont {Zhu},
  \citenamefont {Zhang}, \citenamefont {Pang}, \citenamefont {Qiao},
  \citenamefont {Liu},\ and\ \citenamefont {Wu}}]{zhu2011quantum}%
  \BibitemOpen
  \bibfield  {author} {\bibinfo {author} {\bibfnamefont {X.}~\bibnamefont
  {Zhu}}, \bibinfo {author} {\bibfnamefont {Y.}~\bibnamefont {Zhang}}, \bibinfo
  {author} {\bibfnamefont {S.}~\bibnamefont {Pang}}, \bibinfo {author}
  {\bibfnamefont {C.}~\bibnamefont {Qiao}}, \bibinfo {author} {\bibfnamefont
  {Q.}~\bibnamefont {Liu}},\ and\ \bibinfo {author} {\bibfnamefont
  {S.}~\bibnamefont {Wu}},\ }\bibfield  {title} {\bibinfo {title} {Quantum
  measurements with preselection and postselection},\ }\href
  {https://doi.org/10.1103/PhysRevA.84.052111} {\bibfield  {journal} {\bibinfo
  {journal} {Phys. Rev. A}\ }\textbf {\bibinfo {volume} {84}},\ \bibinfo
  {pages} {052111} (\bibinfo {year} {2011})}\BibitemShut {NoStop}%
\bibitem [{\citenamefont {Elliott}\ \emph {et~al.}(2015)\citenamefont
  {Elliott}, \citenamefont {Kozlowski}, \citenamefont {Caballero-Benitez},\
  and\ \citenamefont {Mekhov}}]{elliott2015multipartite}%
  \BibitemOpen
  \bibfield  {author} {\bibinfo {author} {\bibfnamefont {T.~J.}\ \bibnamefont
  {Elliott}}, \bibinfo {author} {\bibfnamefont {W.}~\bibnamefont {Kozlowski}},
  \bibinfo {author} {\bibfnamefont {S.~F.}\ \bibnamefont {Caballero-Benitez}},\
  and\ \bibinfo {author} {\bibfnamefont {I.~B.}\ \bibnamefont {Mekhov}},\
  }\bibfield  {title} {\bibinfo {title} {Multipartite entangled spatial modes
  of ultracold atoms generated and controlled by quantum measurement},\ }\href
  {https://doi.org/10.1103/PhysRevLett.114.113604} {\bibfield  {journal}
  {\bibinfo  {journal} {Phys. Rev. Lett.}\ }\textbf {\bibinfo {volume} {114}},\
  \bibinfo {pages} {113604} (\bibinfo {year} {2015})}\BibitemShut {NoStop}%
\bibitem [{\citenamefont {Dhar}\ and\ \citenamefont
  {Dasgupta}(2016)}]{dhar2016measurement}%
  \BibitemOpen
  \bibfield  {author} {\bibinfo {author} {\bibfnamefont {S.}~\bibnamefont
  {Dhar}}\ and\ \bibinfo {author} {\bibfnamefont {S.}~\bibnamefont
  {Dasgupta}},\ }\bibfield  {title} {\bibinfo {title} {Measurement-induced
  phase transition in a quantum spin system},\ }\href
  {https://doi.org/10.1103/PhysRevA.93.050103} {\bibfield  {journal} {\bibinfo
  {journal} {Phys. Rev. A}\ }\textbf {\bibinfo {volume} {93}},\ \bibinfo
  {pages} {050103} (\bibinfo {year} {2016})}\BibitemShut {NoStop}%
\bibitem [{\citenamefont {Mazzucchi}\ \emph {et~al.}(2016)\citenamefont
  {Mazzucchi}, \citenamefont {Kozlowski}, \citenamefont {Caballero-Benitez},
  \citenamefont {Elliott},\ and\ \citenamefont
  {Mekhov}}]{mazzucchi2016quantum}%
  \BibitemOpen
  \bibfield  {author} {\bibinfo {author} {\bibfnamefont {G.}~\bibnamefont
  {Mazzucchi}}, \bibinfo {author} {\bibfnamefont {W.}~\bibnamefont
  {Kozlowski}}, \bibinfo {author} {\bibfnamefont {S.~F.}\ \bibnamefont
  {Caballero-Benitez}}, \bibinfo {author} {\bibfnamefont {T.~J.}\ \bibnamefont
  {Elliott}},\ and\ \bibinfo {author} {\bibfnamefont {I.~B.}\ \bibnamefont
  {Mekhov}},\ }\bibfield  {title} {\bibinfo {title} {Quantum
  measurement-induced dynamics of many-body ultracold bosonic and fermionic
  systems in optical lattices},\ }\href
  {https://doi.org/10.1103/PhysRevA.93.023632} {\bibfield  {journal} {\bibinfo
  {journal} {Phys. Rev. A}\ }\textbf {\bibinfo {volume} {93}},\ \bibinfo
  {pages} {023632} (\bibinfo {year} {2016})}\BibitemShut {NoStop}%
\bibitem [{\citenamefont {Li}\ \emph {et~al.}(2018)\citenamefont {Li},
  \citenamefont {Chen},\ and\ \citenamefont {Fisher}}]{li2018quantum}%
  \BibitemOpen
  \bibfield  {author} {\bibinfo {author} {\bibfnamefont {Y.}~\bibnamefont
  {Li}}, \bibinfo {author} {\bibfnamefont {X.}~\bibnamefont {Chen}},\ and\
  \bibinfo {author} {\bibfnamefont {M.~P.~A.}\ \bibnamefont {Fisher}},\
  }\bibfield  {title} {\bibinfo {title} {Quantum {Zeno} effect and the
  many-body entanglement transition},\ }\href
  {https://doi.org/10.1103/PhysRevB.98.205136} {\bibfield  {journal} {\bibinfo
  {journal} {Phys. Rev. B}\ }\textbf {\bibinfo {volume} {98}},\ \bibinfo
  {pages} {205136} (\bibinfo {year} {2018})}\BibitemShut {NoStop}%
\bibitem [{\citenamefont {Chan}\ \emph {et~al.}(2019)\citenamefont {Chan},
  \citenamefont {Nandkishore}, \citenamefont {Pretko},\ and\ \citenamefont
  {Smith}}]{chan2019unitary}%
  \BibitemOpen
  \bibfield  {author} {\bibinfo {author} {\bibfnamefont {A.}~\bibnamefont
  {Chan}}, \bibinfo {author} {\bibfnamefont {R.~M.}\ \bibnamefont
  {Nandkishore}}, \bibinfo {author} {\bibfnamefont {M.}~\bibnamefont
  {Pretko}},\ and\ \bibinfo {author} {\bibfnamefont {G.}~\bibnamefont
  {Smith}},\ }\bibfield  {title} {\bibinfo {title} {Unitary-projective
  entanglement dynamics},\ }\href {https://doi.org/10.1103/PhysRevB.99.224307}
  {\bibfield  {journal} {\bibinfo  {journal} {Phys. Rev. B}\ }\textbf {\bibinfo
  {volume} {99}},\ \bibinfo {pages} {224307} (\bibinfo {year}
  {2019})}\BibitemShut {NoStop}%
\bibitem [{\citenamefont {Li}\ \emph {et~al.}(2019)\citenamefont {Li},
  \citenamefont {Chen},\ and\ \citenamefont {Fisher}}]{li2019measurement}%
  \BibitemOpen
  \bibfield  {author} {\bibinfo {author} {\bibfnamefont {Y.}~\bibnamefont
  {Li}}, \bibinfo {author} {\bibfnamefont {X.}~\bibnamefont {Chen}},\ and\
  \bibinfo {author} {\bibfnamefont {M.~P.}\ \bibnamefont {Fisher}},\ }\bibfield
   {title} {\bibinfo {title} {Measurement-driven entanglement transition in
  hybrid quantum circuits},\ }\href
  {https://doi.org/10.1103/PhysRevB.100.134306} {\bibfield  {journal} {\bibinfo
   {journal} {Phys. Rev. B}\ }\textbf {\bibinfo {volume} {100}},\ \bibinfo
  {pages} {134306} (\bibinfo {year} {2019})}\BibitemShut {NoStop}%
\bibitem [{\citenamefont {Skinner}\ \emph {et~al.}(2019)\citenamefont
  {Skinner}, \citenamefont {Ruhman},\ and\ \citenamefont
  {Nahum}}]{skinner2019measurement}%
  \BibitemOpen
  \bibfield  {author} {\bibinfo {author} {\bibfnamefont {B.}~\bibnamefont
  {Skinner}}, \bibinfo {author} {\bibfnamefont {J.}~\bibnamefont {Ruhman}},\
  and\ \bibinfo {author} {\bibfnamefont {A.}~\bibnamefont {Nahum}},\ }\bibfield
   {title} {\bibinfo {title} {Measurement-induced phase transitions in the
  dynamics of entanglement},\ }\href
  {https://doi.org/10.1103/PhysRevX.9.031009} {\bibfield  {journal} {\bibinfo
  {journal} {Phys. Rev. X}\ }\textbf {\bibinfo {volume} {9}},\ \bibinfo {pages}
  {031009} (\bibinfo {year} {2019})}\BibitemShut {NoStop}%
\bibitem [{\citenamefont {Szyniszewski}\ \emph {et~al.}(2019)\citenamefont
  {Szyniszewski}, \citenamefont {Romito},\ and\ \citenamefont
  {Schomerus}}]{szyniszewski2019entanglement}%
  \BibitemOpen
  \bibfield  {author} {\bibinfo {author} {\bibfnamefont {M.}~\bibnamefont
  {Szyniszewski}}, \bibinfo {author} {\bibfnamefont {A.}~\bibnamefont
  {Romito}},\ and\ \bibinfo {author} {\bibfnamefont {H.}~\bibnamefont
  {Schomerus}},\ }\bibfield  {title} {\bibinfo {title} {Entanglement transition
  from variable-strength weak measurements},\ }\href
  {https://doi.org/10.1103/PhysRevB.100.064204} {\bibfield  {journal} {\bibinfo
   {journal} {Phys. Rev. B}\ }\textbf {\bibinfo {volume} {100}},\ \bibinfo
  {pages} {064204} (\bibinfo {year} {2019})}\BibitemShut {NoStop}%
\bibitem [{\citenamefont {Zabalo}\ \emph {et~al.}(2020)\citenamefont {Zabalo},
  \citenamefont {Gullans}, \citenamefont {Wilson}, \citenamefont
  {Gopalakrishnan}, \citenamefont {Huse},\ and\ \citenamefont
  {Pixley}}]{zabalo2020critical}%
  \BibitemOpen
  \bibfield  {author} {\bibinfo {author} {\bibfnamefont {A.}~\bibnamefont
  {Zabalo}}, \bibinfo {author} {\bibfnamefont {M.~J.}\ \bibnamefont {Gullans}},
  \bibinfo {author} {\bibfnamefont {J.~H.}\ \bibnamefont {Wilson}}, \bibinfo
  {author} {\bibfnamefont {S.}~\bibnamefont {Gopalakrishnan}}, \bibinfo
  {author} {\bibfnamefont {D.~A.}\ \bibnamefont {Huse}},\ and\ \bibinfo
  {author} {\bibfnamefont {J.~H.}\ \bibnamefont {Pixley}},\ }\bibfield  {title}
  {\bibinfo {title} {Critical properties of the measurement-induced transition
  in random quantum circuits},\ }\href
  {https://doi.org/10.1103/PhysRevB.101.060301} {\bibfield  {journal} {\bibinfo
   {journal} {Phys. Rev. B}\ }\textbf {\bibinfo {volume} {101}},\ \bibinfo
  {pages} {060301} (\bibinfo {year} {2020})}\BibitemShut {NoStop}%
\bibitem [{\citenamefont {Nahum}\ \emph {et~al.}(2021)\citenamefont {Nahum},
  \citenamefont {Roy}, \citenamefont {Skinner},\ and\ \citenamefont
  {Ruhman}}]{nahum2021measurement}%
  \BibitemOpen
  \bibfield  {author} {\bibinfo {author} {\bibfnamefont {A.}~\bibnamefont
  {Nahum}}, \bibinfo {author} {\bibfnamefont {S.}~\bibnamefont {Roy}}, \bibinfo
  {author} {\bibfnamefont {B.}~\bibnamefont {Skinner}},\ and\ \bibinfo {author}
  {\bibfnamefont {J.}~\bibnamefont {Ruhman}},\ }\bibfield  {title} {\bibinfo
  {title} {Measurement and entanglement phase transitions in all-to-all quantum
  circuits, on quantum trees, and in landau-ginsburg theory},\ }\href
  {https://doi.org/10.1103/PRXQuantum.2.010352} {\bibfield  {journal} {\bibinfo
   {journal} {PRX Quantum}\ }\textbf {\bibinfo {volume} {2}},\ \bibinfo {pages}
  {010352} (\bibinfo {year} {2021})}\BibitemShut {NoStop}%
\bibitem [{\citenamefont {Tantivasadakarn}\ \emph {et~al.}(2021)\citenamefont
  {Tantivasadakarn}, \citenamefont {Thorngren}, \citenamefont {Vishwanath},\
  and\ \citenamefont {Verresen}}]{tantivasadakarn2021long}%
  \BibitemOpen
  \bibfield  {author} {\bibinfo {author} {\bibfnamefont {N.}~\bibnamefont
  {Tantivasadakarn}}, \bibinfo {author} {\bibfnamefont {R.}~\bibnamefont
  {Thorngren}}, \bibinfo {author} {\bibfnamefont {A.}~\bibnamefont
  {Vishwanath}},\ and\ \bibinfo {author} {\bibfnamefont {R.}~\bibnamefont
  {Verresen}},\ }\bibfield  {title} {\bibinfo {title} {Long-range entanglement
  from measuring symmetry-protected topological phases},\ }\href
  {https://arxiv.org/abs/2112.01519} {\bibfield  {journal} {\bibinfo  {journal}
  {arXiv preprint arXiv:2112.01519}\ } (\bibinfo {year} {2021})}\BibitemShut
  {NoStop}%
\bibitem [{\citenamefont {Sang}\ and\ \citenamefont
  {Hsieh}(2021)}]{sang2021measurement}%
  \BibitemOpen
  \bibfield  {author} {\bibinfo {author} {\bibfnamefont {S.}~\bibnamefont
  {Sang}}\ and\ \bibinfo {author} {\bibfnamefont {T.~H.}\ \bibnamefont
  {Hsieh}},\ }\bibfield  {title} {\bibinfo {title} {Measurement-protected
  quantum phases},\ }\href {https://doi.org/10.1103/PhysRevResearch.3.023200}
  {\bibfield  {journal} {\bibinfo  {journal} {Phys. Rev. Res.}\ }\textbf
  {\bibinfo {volume} {3}},\ \bibinfo {pages} {023200} (\bibinfo {year}
  {2021})}\BibitemShut {NoStop}%
\bibitem [{\citenamefont {Lunt}\ \emph {et~al.}(2021)\citenamefont {Lunt},
  \citenamefont {Szyniszewski},\ and\ \citenamefont
  {Pal}}]{lunt2021measurement}%
  \BibitemOpen
  \bibfield  {author} {\bibinfo {author} {\bibfnamefont {O.}~\bibnamefont
  {Lunt}}, \bibinfo {author} {\bibfnamefont {M.}~\bibnamefont {Szyniszewski}},\
  and\ \bibinfo {author} {\bibfnamefont {A.}~\bibnamefont {Pal}},\ }\bibfield
  {title} {\bibinfo {title} {Measurement-induced criticality and entanglement
  clusters: A study of one-dimensional and two-dimensional clifford circuits},\
  }\href {https://doi.org/10.1103/PhysRevB.104.155111} {\bibfield  {journal}
  {\bibinfo  {journal} {Phys. Rev. B}\ }\textbf {\bibinfo {volume} {104}},\
  \bibinfo {pages} {155111} (\bibinfo {year} {2021})}\BibitemShut {NoStop}%
\bibitem [{\citenamefont {Turkeshi}\ \emph {et~al.}(2020)\citenamefont
  {Turkeshi}, \citenamefont {Fazio},\ and\ \citenamefont
  {Dalmonte}}]{turkeshi2020measurement}%
  \BibitemOpen
  \bibfield  {author} {\bibinfo {author} {\bibfnamefont {X.}~\bibnamefont
  {Turkeshi}}, \bibinfo {author} {\bibfnamefont {R.}~\bibnamefont {Fazio}},\
  and\ \bibinfo {author} {\bibfnamefont {M.}~\bibnamefont {Dalmonte}},\
  }\bibfield  {title} {\bibinfo {title} {Measurement-induced criticality in
  $(2+1)$-dimensional hybrid quantum circuits},\ }\href
  {https://doi.org/10.1103/PhysRevB.102.014315} {\bibfield  {journal} {\bibinfo
   {journal} {Phys. Rev. B}\ }\textbf {\bibinfo {volume} {102}},\ \bibinfo
  {pages} {014315} (\bibinfo {year} {2020})}\BibitemShut {NoStop}%
\bibitem [{\citenamefont {Yu}\ and\ \citenamefont
  {Qi}(2022)}]{yu2022measurement}%
  \BibitemOpen
  \bibfield  {author} {\bibinfo {author} {\bibfnamefont {X.}~\bibnamefont
  {Yu}}\ and\ \bibinfo {author} {\bibfnamefont {X.-L.}\ \bibnamefont {Qi}},\
  }\bibfield  {title} {\bibinfo {title} {Measurement-induced entanglement phase
  transition in random bilocal circuits},\ }\href@noop {} {\bibfield  {journal}
  {\bibinfo  {journal} {arXiv preprint arXiv:2201.12704}\ } (\bibinfo {year}
  {2022})}\BibitemShut {NoStop}%
\bibitem [{\citenamefont {Tang}\ and\ \citenamefont
  {Zhu}(2020)}]{tang2020measurement}%
  \BibitemOpen
  \bibfield  {author} {\bibinfo {author} {\bibfnamefont {Q.}~\bibnamefont
  {Tang}}\ and\ \bibinfo {author} {\bibfnamefont {W.}~\bibnamefont {Zhu}},\
  }\bibfield  {title} {\bibinfo {title} {Measurement-induced phase transition:
  A case study in the nonintegrable model by density-matrix renormalization
  group calculations},\ }\href
  {https://doi.org/10.1103/PhysRevResearch.2.013022} {\bibfield  {journal}
  {\bibinfo  {journal} {Phys. Rev. Res.}\ }\textbf {\bibinfo {volume} {2}},\
  \bibinfo {pages} {013022} (\bibinfo {year} {2020})}\BibitemShut {NoStop}%
\bibitem [{\citenamefont {Turkeshi}\ \emph {et~al.}(2021)\citenamefont
  {Turkeshi}, \citenamefont {Biella}, \citenamefont {Fazio}, \citenamefont
  {Dalmonte},\ and\ \citenamefont {Schir\'o}}]{turkeshi2021measurement}%
  \BibitemOpen
  \bibfield  {author} {\bibinfo {author} {\bibfnamefont {X.}~\bibnamefont
  {Turkeshi}}, \bibinfo {author} {\bibfnamefont {A.}~\bibnamefont {Biella}},
  \bibinfo {author} {\bibfnamefont {R.}~\bibnamefont {Fazio}}, \bibinfo
  {author} {\bibfnamefont {M.}~\bibnamefont {Dalmonte}},\ and\ \bibinfo
  {author} {\bibfnamefont {M.}~\bibnamefont {Schir\'o}},\ }\bibfield  {title}
  {\bibinfo {title} {Measurement-induced entanglement transitions in the
  quantum {Ising} chain: From infinite to zero clicks},\ }\href
  {https://doi.org/10.1103/PhysRevB.103.224210} {\bibfield  {journal} {\bibinfo
   {journal} {Phys. Rev. B}\ }\textbf {\bibinfo {volume} {103}},\ \bibinfo
  {pages} {224210} (\bibinfo {year} {2021})}\BibitemShut {NoStop}%
\bibitem [{\citenamefont {Lavasani}\ \emph {et~al.}(2021)\citenamefont
  {Lavasani}, \citenamefont {Alavirad},\ and\ \citenamefont
  {Barkeshli}}]{lavasani2021measurement}%
  \BibitemOpen
  \bibfield  {author} {\bibinfo {author} {\bibfnamefont {A.}~\bibnamefont
  {Lavasani}}, \bibinfo {author} {\bibfnamefont {Y.}~\bibnamefont {Alavirad}},\
  and\ \bibinfo {author} {\bibfnamefont {M.}~\bibnamefont {Barkeshli}},\
  }\bibfield  {title} {\bibinfo {title} {Measurement-induced topological
  entanglement transitions in symmetric random quantum circuits},\ }\href
  {https://doi.org/10.1038/s41567-020-01112-z} {\bibfield  {journal} {\bibinfo
  {journal} {Nat. Phys.}\ }\textbf {\bibinfo {volume} {17}},\ \bibinfo {pages}
  {342} (\bibinfo {year} {2021})}\BibitemShut {NoStop}%
\bibitem [{\citenamefont {Choi}\ \emph {et~al.}(2020)\citenamefont {Choi},
  \citenamefont {Bao}, \citenamefont {Qi},\ and\ \citenamefont
  {Altman}}]{choi2020quantum}%
  \BibitemOpen
  \bibfield  {author} {\bibinfo {author} {\bibfnamefont {S.}~\bibnamefont
  {Choi}}, \bibinfo {author} {\bibfnamefont {Y.}~\bibnamefont {Bao}}, \bibinfo
  {author} {\bibfnamefont {X.-L.}\ \bibnamefont {Qi}},\ and\ \bibinfo {author}
  {\bibfnamefont {E.}~\bibnamefont {Altman}},\ }\bibfield  {title} {\bibinfo
  {title} {Quantum error correction in scrambling dynamics and
  measurement-induced phase transition},\ }\href
  {https://doi.org/10.1103/PhysRevLett.125.030505} {\bibfield  {journal}
  {\bibinfo  {journal} {Phys. Rev. Lett.}\ }\textbf {\bibinfo {volume} {125}},\
  \bibinfo {pages} {030505} (\bibinfo {year} {2020})}\BibitemShut {NoStop}%
\bibitem [{\citenamefont {Li}\ and\ \citenamefont
  {Fisher}(2021)}]{li2021statistical}%
  \BibitemOpen
  \bibfield  {author} {\bibinfo {author} {\bibfnamefont {Y.}~\bibnamefont
  {Li}}\ and\ \bibinfo {author} {\bibfnamefont {M.~P.~A.}\ \bibnamefont
  {Fisher}},\ }\bibfield  {title} {\bibinfo {title} {Statistical mechanics of
  quantum error correcting codes},\ }\href
  {https://doi.org/10.1103/PhysRevB.103.104306} {\bibfield  {journal} {\bibinfo
   {journal} {Phys. Rev. B}\ }\textbf {\bibinfo {volume} {103}},\ \bibinfo
  {pages} {104306} (\bibinfo {year} {2021})}\BibitemShut {NoStop}%
\bibitem [{\citenamefont {Fan}\ \emph {et~al.}(2021)\citenamefont {Fan},
  \citenamefont {Vijay}, \citenamefont {Vishwanath},\ and\ \citenamefont
  {You}}]{fan2021self}%
  \BibitemOpen
  \bibfield  {author} {\bibinfo {author} {\bibfnamefont {R.}~\bibnamefont
  {Fan}}, \bibinfo {author} {\bibfnamefont {S.}~\bibnamefont {Vijay}}, \bibinfo
  {author} {\bibfnamefont {A.}~\bibnamefont {Vishwanath}},\ and\ \bibinfo
  {author} {\bibfnamefont {Y.-Z.}\ \bibnamefont {You}},\ }\bibfield  {title}
  {\bibinfo {title} {Self-organized error correction in random unitary circuits
  with measurement},\ }\href {https://doi.org/10.1103/PhysRevB.103.174309}
  {\bibfield  {journal} {\bibinfo  {journal} {Phys. Rev. B}\ }\textbf {\bibinfo
  {volume} {103}},\ \bibinfo {pages} {174309} (\bibinfo {year}
  {2021})}\BibitemShut {NoStop}%
\bibitem [{\citenamefont {Vasseur}\ \emph {et~al.}(2019)\citenamefont
  {Vasseur}, \citenamefont {Potter}, \citenamefont {You},\ and\ \citenamefont
  {Ludwig}}]{vasseur2019entangling}%
  \BibitemOpen
  \bibfield  {author} {\bibinfo {author} {\bibfnamefont {R.}~\bibnamefont
  {Vasseur}}, \bibinfo {author} {\bibfnamefont {A.~C.}\ \bibnamefont {Potter}},
  \bibinfo {author} {\bibfnamefont {Y.-Z.}\ \bibnamefont {You}},\ and\ \bibinfo
  {author} {\bibfnamefont {A.~W.~W.}\ \bibnamefont {Ludwig}},\ }\bibfield
  {title} {\bibinfo {title} {Entanglement transitions from holographic random
  tensor networks},\ }\href {https://doi.org/10.1103/PhysRevB.100.134203}
  {\bibfield  {journal} {\bibinfo  {journal} {Phys. Rev. B}\ }\textbf {\bibinfo
  {volume} {100}},\ \bibinfo {pages} {134203} (\bibinfo {year}
  {2019})}\BibitemShut {NoStop}%
\bibitem [{\citenamefont {Bao}\ \emph {et~al.}(2020)\citenamefont {Bao},
  \citenamefont {Choi},\ and\ \citenamefont {Altman}}]{bao2020theory}%
  \BibitemOpen
  \bibfield  {author} {\bibinfo {author} {\bibfnamefont {Y.}~\bibnamefont
  {Bao}}, \bibinfo {author} {\bibfnamefont {S.}~\bibnamefont {Choi}},\ and\
  \bibinfo {author} {\bibfnamefont {E.}~\bibnamefont {Altman}},\ }\bibfield
  {title} {\bibinfo {title} {Theory of the phase transition in random unitary
  circuits with measurements},\ }\href
  {https://doi.org/10.1103/PhysRevB.101.104301} {\bibfield  {journal} {\bibinfo
   {journal} {Phys. Rev. B}\ }\textbf {\bibinfo {volume} {101}},\ \bibinfo
  {pages} {104301} (\bibinfo {year} {2020})}\BibitemShut {NoStop}%
\bibitem [{\citenamefont {Jian}\ \emph {et~al.}(2020)\citenamefont {Jian},
  \citenamefont {You}, \citenamefont {Vasseur},\ and\ \citenamefont
  {Ludwig}}]{jian2020measurement}%
  \BibitemOpen
  \bibfield  {author} {\bibinfo {author} {\bibfnamefont {C.-M.}\ \bibnamefont
  {Jian}}, \bibinfo {author} {\bibfnamefont {Y.-Z.}\ \bibnamefont {You}},
  \bibinfo {author} {\bibfnamefont {R.}~\bibnamefont {Vasseur}},\ and\ \bibinfo
  {author} {\bibfnamefont {A.~W.~W.}\ \bibnamefont {Ludwig}},\ }\bibfield
  {title} {\bibinfo {title} {Measurement-induced criticality in random quantum
  circuits},\ }\href {https://doi.org/10.1103/PhysRevB.101.104302} {\bibfield
  {journal} {\bibinfo  {journal} {Phys. Rev. B}\ }\textbf {\bibinfo {volume}
  {101}},\ \bibinfo {pages} {104302} (\bibinfo {year} {2020})}\BibitemShut
  {NoStop}%
\bibitem [{\citenamefont {Sang}\ \emph {et~al.}(2021)\citenamefont {Sang},
  \citenamefont {Li}, \citenamefont {Zhou}, \citenamefont {Chen}, \citenamefont
  {Hsieh},\ and\ \citenamefont {Fisher}}]{sang2021entanglement}%
  \BibitemOpen
  \bibfield  {author} {\bibinfo {author} {\bibfnamefont {S.}~\bibnamefont
  {Sang}}, \bibinfo {author} {\bibfnamefont {Y.}~\bibnamefont {Li}}, \bibinfo
  {author} {\bibfnamefont {T.}~\bibnamefont {Zhou}}, \bibinfo {author}
  {\bibfnamefont {X.}~\bibnamefont {Chen}}, \bibinfo {author} {\bibfnamefont
  {T.~H.}\ \bibnamefont {Hsieh}},\ and\ \bibinfo {author} {\bibfnamefont
  {M.~P.}\ \bibnamefont {Fisher}},\ }\bibfield  {title} {\bibinfo {title}
  {Entanglement negativity at measurement-induced criticality},\ }\href
  {https://doi.org/10.1103/PRXQuantum.2.030313} {\bibfield  {journal} {\bibinfo
   {journal} {PRX Quantum}\ }\textbf {\bibinfo {volume} {2}},\ \bibinfo {pages}
  {030313} (\bibinfo {year} {2021})}\BibitemShut {NoStop}%
\bibitem [{\citenamefont {Block}\ \emph {et~al.}(2022)\citenamefont {Block},
  \citenamefont {Bao}, \citenamefont {Choi}, \citenamefont {Altman},\ and\
  \citenamefont {Yao}}]{block2022measurement}%
  \BibitemOpen
  \bibfield  {author} {\bibinfo {author} {\bibfnamefont {M.}~\bibnamefont
  {Block}}, \bibinfo {author} {\bibfnamefont {Y.}~\bibnamefont {Bao}}, \bibinfo
  {author} {\bibfnamefont {S.}~\bibnamefont {Choi}}, \bibinfo {author}
  {\bibfnamefont {E.}~\bibnamefont {Altman}},\ and\ \bibinfo {author}
  {\bibfnamefont {N.~Y.}\ \bibnamefont {Yao}},\ }\bibfield  {title} {\bibinfo
  {title} {Measurement-induced transition in long-range interacting quantum
  circuits},\ }\href {https://doi.org/10.1103/PhysRevLett.128.010604}
  {\bibfield  {journal} {\bibinfo  {journal} {Phys. Rev. Lett.}\ }\textbf
  {\bibinfo {volume} {128}},\ \bibinfo {pages} {010604} (\bibinfo {year}
  {2022})}\BibitemShut {NoStop}%
\bibitem [{\citenamefont {Gullans}\ and\ \citenamefont
  {Huse}(2020{\natexlab{a}})}]{gullans2020dynamical}%
  \BibitemOpen
  \bibfield  {author} {\bibinfo {author} {\bibfnamefont {M.~J.}\ \bibnamefont
  {Gullans}}\ and\ \bibinfo {author} {\bibfnamefont {D.~A.}\ \bibnamefont
  {Huse}},\ }\bibfield  {title} {\bibinfo {title} {Dynamical purification phase
  transition induced by quantum measurements},\ }\href
  {https://doi.org/10.1103/PhysRevX.10.041020} {\bibfield  {journal} {\bibinfo
  {journal} {Phys. Rev. X}\ }\textbf {\bibinfo {volume} {10}},\ \bibinfo
  {pages} {041020} (\bibinfo {year} {2020}{\natexlab{a}})}\BibitemShut
  {NoStop}%
\bibitem [{\citenamefont {Noel}\ \emph {et~al.}(2021)\citenamefont {Noel},
  \citenamefont {Niroula}, \citenamefont {Zhu}, \citenamefont {Risinger},
  \citenamefont {Egan}, \citenamefont {Biswas}, \citenamefont {Cetina},
  \citenamefont {Gorshkov}, \citenamefont {Gullans}, \citenamefont {Huse},\
  and\ \citenamefont {Monroe}}]{noel2021observation}%
  \BibitemOpen
  \bibfield  {author} {\bibinfo {author} {\bibfnamefont {C.}~\bibnamefont
  {Noel}}, \bibinfo {author} {\bibfnamefont {P.}~\bibnamefont {Niroula}},
  \bibinfo {author} {\bibfnamefont {D.}~\bibnamefont {Zhu}}, \bibinfo {author}
  {\bibfnamefont {A.}~\bibnamefont {Risinger}}, \bibinfo {author}
  {\bibfnamefont {L.}~\bibnamefont {Egan}}, \bibinfo {author} {\bibfnamefont
  {D.}~\bibnamefont {Biswas}}, \bibinfo {author} {\bibfnamefont
  {M.}~\bibnamefont {Cetina}}, \bibinfo {author} {\bibfnamefont {A.~V.}\
  \bibnamefont {Gorshkov}}, \bibinfo {author} {\bibfnamefont {M.~J.}\
  \bibnamefont {Gullans}}, \bibinfo {author} {\bibfnamefont {D.~A.}\
  \bibnamefont {Huse}},\ and\ \bibinfo {author} {\bibfnamefont
  {C.}~\bibnamefont {Monroe}},\ }\href {https://arxiv.org/abs/2106.05881}
  {\bibinfo {title} {Observation of measurement-induced quantum phases in a
  trapped-ion quantum computer}} (\bibinfo {year} {2021}),\ \Eprint
  {https://arxiv.org/abs/2106.05881} {arXiv:2106.05881 [quant-ph]} \BibitemShut
  {NoStop}%
\bibitem [{\citenamefont {C\'orcoles}\ \emph {et~al.}(2021)\citenamefont
  {C\'orcoles}, \citenamefont {Takita}, \citenamefont {Inoue}, \citenamefont
  {Lekuch}, \citenamefont {Minev}, \citenamefont {Chow},\ and\ \citenamefont
  {Gambetta}}]{corcoles2021exploiting}%
  \BibitemOpen
  \bibfield  {author} {\bibinfo {author} {\bibfnamefont {A.~D.}\ \bibnamefont
  {C\'orcoles}}, \bibinfo {author} {\bibfnamefont {M.}~\bibnamefont {Takita}},
  \bibinfo {author} {\bibfnamefont {K.}~\bibnamefont {Inoue}}, \bibinfo
  {author} {\bibfnamefont {S.}~\bibnamefont {Lekuch}}, \bibinfo {author}
  {\bibfnamefont {Z.~K.}\ \bibnamefont {Minev}}, \bibinfo {author}
  {\bibfnamefont {J.~M.}\ \bibnamefont {Chow}},\ and\ \bibinfo {author}
  {\bibfnamefont {J.~M.}\ \bibnamefont {Gambetta}},\ }\bibfield  {title}
  {\bibinfo {title} {Exploiting dynamic quantum circuits in a quantum algorithm
  with superconducting qubits},\ }\href
  {https://doi.org/10.1103/PhysRevLett.127.100501} {\bibfield  {journal}
  {\bibinfo  {journal} {Phys. Rev. Lett.}\ }\textbf {\bibinfo {volume} {127}},\
  \bibinfo {pages} {100501} (\bibinfo {year} {2021})}\BibitemShut {NoStop}%
\bibitem [{\citenamefont {Gebhart}\ \emph {et~al.}(2020)\citenamefont
  {Gebhart}, \citenamefont {Snizhko}, \citenamefont {Wellens}, \citenamefont
  {Buchleitner}, \citenamefont {Romito},\ and\ \citenamefont
  {Gefen}}]{gebhart2020topological}%
  \BibitemOpen
  \bibfield  {author} {\bibinfo {author} {\bibfnamefont {V.}~\bibnamefont
  {Gebhart}}, \bibinfo {author} {\bibfnamefont {K.}~\bibnamefont {Snizhko}},
  \bibinfo {author} {\bibfnamefont {T.}~\bibnamefont {Wellens}}, \bibinfo
  {author} {\bibfnamefont {A.}~\bibnamefont {Buchleitner}}, \bibinfo {author}
  {\bibfnamefont {A.}~\bibnamefont {Romito}},\ and\ \bibinfo {author}
  {\bibfnamefont {Y.}~\bibnamefont {Gefen}},\ }\bibfield  {title} {\bibinfo
  {title} {Topological transition in measurement-induced geometric phases},\
  }\href {https://doi.org/10.1073/pnas.1911620117} {\bibfield  {journal}
  {\bibinfo  {journal} {Proc. Natl. Acad. Sci. U.S.A.}\ }\textbf {\bibinfo
  {volume} {117}},\ \bibinfo {pages} {5706} (\bibinfo {year}
  {2020})}\BibitemShut {NoStop}%
\bibitem [{\citenamefont {Zilberberg}\ \emph {et~al.}(2013)\citenamefont
  {Zilberberg}, \citenamefont {Romito}, \citenamefont {Starling}, \citenamefont
  {Howland}, \citenamefont {Broadbent}, \citenamefont {Howell},\ and\
  \citenamefont {Gefen}}]{zilberberg2013null}%
  \BibitemOpen
  \bibfield  {author} {\bibinfo {author} {\bibfnamefont {O.}~\bibnamefont
  {Zilberberg}}, \bibinfo {author} {\bibfnamefont {A.}~\bibnamefont {Romito}},
  \bibinfo {author} {\bibfnamefont {D.~J.}\ \bibnamefont {Starling}}, \bibinfo
  {author} {\bibfnamefont {G.~A.}\ \bibnamefont {Howland}}, \bibinfo {author}
  {\bibfnamefont {C.~J.}\ \bibnamefont {Broadbent}}, \bibinfo {author}
  {\bibfnamefont {J.~C.}\ \bibnamefont {Howell}},\ and\ \bibinfo {author}
  {\bibfnamefont {Y.}~\bibnamefont {Gefen}},\ }\bibfield  {title} {\bibinfo
  {title} {Null values and quantum state discr\`{\i}mination},\ }\href
  {https://doi.org/10.1103/PhysRevLett.110.170405} {\bibfield  {journal}
  {\bibinfo  {journal} {Phys. Rev. Lett.}\ }\textbf {\bibinfo {volume} {110}},\
  \bibinfo {pages} {170405} (\bibinfo {year} {2013})}\BibitemShut {NoStop}%
\bibitem [{\citenamefont {Jurcevic}\ \emph {et~al.}(2021)\citenamefont
  {Jurcevic}, \citenamefont {Javadi-Abhari}, \citenamefont {Bishop},
  \citenamefont {Lauer}, \citenamefont {Bogorin}, \citenamefont {Brink},
  \citenamefont {Capelluto}, \citenamefont {G{\"u}nl{\"u}k}, \citenamefont
  {Itoko}, \citenamefont {Kanazawa} \emph
  {et~al.}}]{jurcevic2021demonstration}%
  \BibitemOpen
  \bibfield  {author} {\bibinfo {author} {\bibfnamefont {P.}~\bibnamefont
  {Jurcevic}}, \bibinfo {author} {\bibfnamefont {A.}~\bibnamefont
  {Javadi-Abhari}}, \bibinfo {author} {\bibfnamefont {L.~S.}\ \bibnamefont
  {Bishop}}, \bibinfo {author} {\bibfnamefont {I.}~\bibnamefont {Lauer}},
  \bibinfo {author} {\bibfnamefont {D.~F.}\ \bibnamefont {Bogorin}}, \bibinfo
  {author} {\bibfnamefont {M.}~\bibnamefont {Brink}}, \bibinfo {author}
  {\bibfnamefont {L.}~\bibnamefont {Capelluto}}, \bibinfo {author}
  {\bibfnamefont {O.}~\bibnamefont {G{\"u}nl{\"u}k}}, \bibinfo {author}
  {\bibfnamefont {T.}~\bibnamefont {Itoko}}, \bibinfo {author} {\bibfnamefont
  {N.}~\bibnamefont {Kanazawa}}, \emph {et~al.},\ }\bibfield  {title} {\bibinfo
  {title} {Demonstration of quantum volume 64 on a superconducting quantum
  computing system},\ }\href {https://doi.org/10.1088/2058-9565/abe519}
  {\bibfield  {journal} {\bibinfo  {journal} {Quantum Sci. Technol.}\ }\textbf
  {\bibinfo {volume} {6}},\ \bibinfo {pages} {025020} (\bibinfo {year}
  {2021})}\BibitemShut {NoStop}%
\bibitem [{\citenamefont {Kandala}\ \emph {et~al.}(2017)\citenamefont
  {Kandala}, \citenamefont {Mezzacapo}, \citenamefont {Temme}, \citenamefont
  {Takita}, \citenamefont {Brink}, \citenamefont {Chow},\ and\ \citenamefont
  {Gambetta}}]{kandala2017hardware}%
  \BibitemOpen
  \bibfield  {author} {\bibinfo {author} {\bibfnamefont {A.}~\bibnamefont
  {Kandala}}, \bibinfo {author} {\bibfnamefont {A.}~\bibnamefont {Mezzacapo}},
  \bibinfo {author} {\bibfnamefont {K.}~\bibnamefont {Temme}}, \bibinfo
  {author} {\bibfnamefont {M.}~\bibnamefont {Takita}}, \bibinfo {author}
  {\bibfnamefont {M.}~\bibnamefont {Brink}}, \bibinfo {author} {\bibfnamefont
  {J.~M.}\ \bibnamefont {Chow}},\ and\ \bibinfo {author} {\bibfnamefont
  {J.~M.}\ \bibnamefont {Gambetta}},\ }\bibfield  {title} {\bibinfo {title}
  {Hardware-efficient variational quantum eigensolver for small molecules and
  quantum magnets},\ }\href {https://doi.org/10.1038/nature23879} {\bibfield
  {journal} {\bibinfo  {journal} {Nature}\ }\textbf {\bibinfo {volume} {549}},\
  \bibinfo {pages} {242} (\bibinfo {year} {2017})}\BibitemShut {NoStop}%
\bibitem [{\citenamefont {Li}\ \emph {et~al.}(2021)\citenamefont {Li},
  \citenamefont {Chen}, \citenamefont {Ludwig},\ and\ \citenamefont
  {Fisher}}]{li2021conformal}%
  \BibitemOpen
  \bibfield  {author} {\bibinfo {author} {\bibfnamefont {Y.}~\bibnamefont
  {Li}}, \bibinfo {author} {\bibfnamefont {X.}~\bibnamefont {Chen}}, \bibinfo
  {author} {\bibfnamefont {A.~W.~W.}\ \bibnamefont {Ludwig}},\ and\ \bibinfo
  {author} {\bibfnamefont {M.~P.~A.}\ \bibnamefont {Fisher}},\ }\bibfield
  {title} {\bibinfo {title} {Conformal invariance and quantum nonlocality in
  critical hybrid circuits},\ }\href
  {https://doi.org/10.1103/PhysRevB.104.104305} {\bibfield  {journal} {\bibinfo
   {journal} {Phys. Rev. B}\ }\textbf {\bibinfo {volume} {104}},\ \bibinfo
  {pages} {104305} (\bibinfo {year} {2021})}\BibitemShut {NoStop}%
\bibitem [{\citenamefont {Fisher}\ and\ \citenamefont
  {Barber}(1972)}]{fisher1972scaling}%
  \BibitemOpen
  \bibfield  {author} {\bibinfo {author} {\bibfnamefont {M.~E.}\ \bibnamefont
  {Fisher}}\ and\ \bibinfo {author} {\bibfnamefont {M.~N.}\ \bibnamefont
  {Barber}},\ }\bibfield  {title} {\bibinfo {title} {Scaling theory for
  finite-size effects in the critical region},\ }\href
  {https://doi.org/10.1103/PhysRevLett.28.1516} {\bibfield  {journal} {\bibinfo
   {journal} {Phys. Rev. Lett.}\ }\textbf {\bibinfo {volume} {28}},\ \bibinfo
  {pages} {1516} (\bibinfo {year} {1972})}\BibitemShut {NoStop}%
\bibitem [{\citenamefont {Binder}\ and\ \citenamefont
  {Heermann}(2010)}]{binder2010monte}%
  \BibitemOpen
  \bibfield  {author} {\bibinfo {author} {\bibfnamefont {K.}~\bibnamefont
  {Binder}}\ and\ \bibinfo {author} {\bibfnamefont {D.~W.}\ \bibnamefont
  {Heermann}},\ }\href {https://doi.org/10.1007/978-3-642-03163-2} {\emph
  {\bibinfo {title} {Monte Carlo Simulation in Statistical Physics}}},\
  \bibinfo {edition} {5th}\ ed.\ (\bibinfo  {publisher} {Springer-Verlag Berlin
  Heidelberg},\ \bibinfo {address} {Heidelberg},\ \bibinfo {year}
  {2010})\BibitemShut {NoStop}%
\bibitem [{\citenamefont {Gullans}\ and\ \citenamefont
  {Huse}(2020{\natexlab{b}})}]{gullans2020scalable}%
  \BibitemOpen
  \bibfield  {author} {\bibinfo {author} {\bibfnamefont {M.~J.}\ \bibnamefont
  {Gullans}}\ and\ \bibinfo {author} {\bibfnamefont {D.~A.}\ \bibnamefont
  {Huse}},\ }\bibfield  {title} {\bibinfo {title} {Scalable probes of
  measurement-induced criticality},\ }\href
  {https://doi.org/10.1103/PhysRevLett.125.070606} {\bibfield  {journal}
  {\bibinfo  {journal} {Phys. Rev. Lett.}\ }\textbf {\bibinfo {volume} {125}},\
  \bibinfo {pages} {070606} (\bibinfo {year} {2020}{\natexlab{b}})}\BibitemShut
  {NoStop}%
\bibitem [{\citenamefont {Nielsen}\ \emph {et~al.}(2021)\citenamefont
  {Nielsen}, \citenamefont {Gamble}, \citenamefont {Rudinger}, \citenamefont
  {Scholten}, \citenamefont {Young},\ and\ \citenamefont
  {Blume-Kohout}}]{nielsen2021gate}%
  \BibitemOpen
  \bibfield  {author} {\bibinfo {author} {\bibfnamefont {E.}~\bibnamefont
  {Nielsen}}, \bibinfo {author} {\bibfnamefont {J.~K.}\ \bibnamefont {Gamble}},
  \bibinfo {author} {\bibfnamefont {K.}~\bibnamefont {Rudinger}}, \bibinfo
  {author} {\bibfnamefont {T.}~\bibnamefont {Scholten}}, \bibinfo {author}
  {\bibfnamefont {K.}~\bibnamefont {Young}},\ and\ \bibinfo {author}
  {\bibfnamefont {R.}~\bibnamefont {Blume-Kohout}},\ }\bibfield  {title}
  {\bibinfo {title} {Gate set tomography},\ }\href
  {https://doi.org/10.22331/q-2021-10-05-557} {\bibfield  {journal} {\bibinfo
  {journal} {Quantum}\ }\textbf {\bibinfo {volume} {5}},\ \bibinfo {pages}
  {557} (\bibinfo {year} {2021})}\BibitemShut {NoStop}%
\bibitem [{\citenamefont {de~Burgh}\ \emph {et~al.}(2008)\citenamefont
  {de~Burgh}, \citenamefont {Langford}, \citenamefont {Doherty},\ and\
  \citenamefont {Gilchrist}}]{de2008choice}%
  \BibitemOpen
  \bibfield  {author} {\bibinfo {author} {\bibfnamefont {M.~D.}\ \bibnamefont
  {de~Burgh}}, \bibinfo {author} {\bibfnamefont {N.~K.}\ \bibnamefont
  {Langford}}, \bibinfo {author} {\bibfnamefont {A.~C.}\ \bibnamefont
  {Doherty}},\ and\ \bibinfo {author} {\bibfnamefont {A.}~\bibnamefont
  {Gilchrist}},\ }\bibfield  {title} {\bibinfo {title} {Choice of measurement
  sets in qubit tomography},\ }\href
  {https://doi.org/10.1103/PhysRevA.78.052122} {\bibfield  {journal} {\bibinfo
  {journal} {Phys. Rev. A}\ }\textbf {\bibinfo {volume} {78}},\ \bibinfo
  {pages} {052122} (\bibinfo {year} {2008})}\BibitemShut {NoStop}%
\bibitem [{\citenamefont {Aleksandrowicz}\ \emph {et~al.}(2019)\citenamefont
  {Aleksandrowicz} \emph {et~al.}}]{Qiskit}%
  \BibitemOpen
  \bibfield  {author} {\bibinfo {author} {\bibfnamefont {G.}~\bibnamefont
  {Aleksandrowicz}} \emph {et~al.},\ }\bibfield  {title} {\bibinfo {title}
  {{Qiskit: An Open-source Framework for Quantum Computing}},\ }\bibfield
  {journal} {\bibinfo  {journal} {Zenodo}\ }\href
  {https://doi.org/10.5281/ZENODO.2562111} {10.5281/ZENODO.2562111} (\bibinfo
  {year} {2019})\BibitemShut {NoStop}%
\bibitem [{\citenamefont {Smolin}\ \emph {et~al.}(2012)\citenamefont {Smolin},
  \citenamefont {Gambetta},\ and\ \citenamefont {Smith}}]{smolin2012efficient}%
  \BibitemOpen
  \bibfield  {author} {\bibinfo {author} {\bibfnamefont {J.~A.}\ \bibnamefont
  {Smolin}}, \bibinfo {author} {\bibfnamefont {J.~M.}\ \bibnamefont
  {Gambetta}},\ and\ \bibinfo {author} {\bibfnamefont {G.}~\bibnamefont
  {Smith}},\ }\bibfield  {title} {\bibinfo {title} {Efficient method for
  computing the maximum-likelihood quantum state from measurements with
  additive {Gaussian} noise},\ }\href
  {https://doi.org/10.1103/PhysRevLett.108.070502} {\bibfield  {journal}
  {\bibinfo  {journal} {Phys. Rev. Lett.}\ }\textbf {\bibinfo {volume} {108}},\
  \bibinfo {pages} {070502} (\bibinfo {year} {2012})}\BibitemShut {NoStop}%
\bibitem [{\citenamefont {Hamamura}\ and\ \citenamefont
  {Imamichi}(2020)}]{hamamura2020efficient}%
  \BibitemOpen
  \bibfield  {author} {\bibinfo {author} {\bibfnamefont {I.}~\bibnamefont
  {Hamamura}}\ and\ \bibinfo {author} {\bibfnamefont {T.}~\bibnamefont
  {Imamichi}},\ }\bibfield  {title} {\bibinfo {title} {Efficient evaluation of
  quantum observables using entangled measurements},\ }\href
  {https://doi.org/10.1038/s41534-020-0284-2} {\bibfield  {journal} {\bibinfo
  {journal} {npj Quantum Inf.}\ }\textbf {\bibinfo {volume} {6}},\ \bibinfo
  {pages} {1} (\bibinfo {year} {2020})}\BibitemShut {NoStop}%
\bibitem [{\citenamefont {Yen}\ \emph {et~al.}(2020)\citenamefont {Yen},
  \citenamefont {Verteletskyi},\ and\ \citenamefont
  {Izmaylov}}]{yen2020measuring}%
  \BibitemOpen
  \bibfield  {author} {\bibinfo {author} {\bibfnamefont {T.-C.}\ \bibnamefont
  {Yen}}, \bibinfo {author} {\bibfnamefont {V.}~\bibnamefont {Verteletskyi}},\
  and\ \bibinfo {author} {\bibfnamefont {A.~F.}\ \bibnamefont {Izmaylov}},\
  }\bibfield  {title} {\bibinfo {title} {Measuring all compatible operators in
  one series of single-qubit measurements using unitary transformations},\
  }\href {https://doi.org/10.1021/acs.jctc.0c00008} {\bibfield  {journal}
  {\bibinfo  {journal} {J. Chem. Theory Comput.}\ }\textbf {\bibinfo {volume}
  {16}},\ \bibinfo {pages} {2400} (\bibinfo {year} {2020})}\BibitemShut
  {NoStop}%
\bibitem [{\citenamefont {Durt}\ \emph {et~al.}(2010)\citenamefont {Durt},
  \citenamefont {Englert}, \citenamefont {Bengtsson},\ and\ \citenamefont
  {{\.Z}yczkowski}}]{durt2010mutually}%
  \BibitemOpen
  \bibfield  {author} {\bibinfo {author} {\bibfnamefont {T.}~\bibnamefont
  {Durt}}, \bibinfo {author} {\bibfnamefont {B.-G.}\ \bibnamefont {Englert}},
  \bibinfo {author} {\bibfnamefont {I.}~\bibnamefont {Bengtsson}},\ and\
  \bibinfo {author} {\bibfnamefont {K.}~\bibnamefont {{\.Z}yczkowski}},\
  }\bibfield  {title} {\bibinfo {title} {On mutually unbiased bases},\ }\href
  {https://doi.org/10.1142/S0219749910006502} {\bibfield  {journal} {\bibinfo
  {journal} {Int. J. Quantum Inf.}\ }\textbf {\bibinfo {volume} {8}},\ \bibinfo
  {pages} {535} (\bibinfo {year} {2010})}\BibitemShut {NoStop}%
\bibitem [{\citenamefont {Romero}\ \emph {et~al.}(2005)\citenamefont {Romero},
  \citenamefont {Bj\"ork}, \citenamefont {Klimov},\ and\ \citenamefont
  {S\'anchez-Soto}}]{romero2005structure}%
  \BibitemOpen
  \bibfield  {author} {\bibinfo {author} {\bibfnamefont {J.~L.}\ \bibnamefont
  {Romero}}, \bibinfo {author} {\bibfnamefont {G.}~\bibnamefont {Bj\"ork}},
  \bibinfo {author} {\bibfnamefont {A.~B.}\ \bibnamefont {Klimov}},\ and\
  \bibinfo {author} {\bibfnamefont {L.~L.}\ \bibnamefont {S\'anchez-Soto}},\
  }\bibfield  {title} {\bibinfo {title} {Structure of the sets of mutually
  unbiased bases for $n$ qubits},\ }\href
  {https://doi.org/10.1103/PhysRevA.72.062310} {\bibfield  {journal} {\bibinfo
  {journal} {Phys. Rev. A}\ }\textbf {\bibinfo {volume} {72}},\ \bibinfo
  {pages} {062310} (\bibinfo {year} {2005})}\BibitemShut {NoStop}%
\bibitem [{\citenamefont {Gokhale}\ \emph {et~al.}(2019)\citenamefont
  {Gokhale}, \citenamefont {Angiuli}, \citenamefont {Ding}, \citenamefont
  {Gui}, \citenamefont {Tomesh}, \citenamefont {Suchara}, \citenamefont
  {Martonosi},\ and\ \citenamefont {Chong}}]{gokhale2019minimizing}%
  \BibitemOpen
  \bibfield  {author} {\bibinfo {author} {\bibfnamefont {P.}~\bibnamefont
  {Gokhale}}, \bibinfo {author} {\bibfnamefont {O.}~\bibnamefont {Angiuli}},
  \bibinfo {author} {\bibfnamefont {Y.}~\bibnamefont {Ding}}, \bibinfo {author}
  {\bibfnamefont {K.}~\bibnamefont {Gui}}, \bibinfo {author} {\bibfnamefont
  {T.}~\bibnamefont {Tomesh}}, \bibinfo {author} {\bibfnamefont
  {M.}~\bibnamefont {Suchara}}, \bibinfo {author} {\bibfnamefont
  {M.}~\bibnamefont {Martonosi}},\ and\ \bibinfo {author} {\bibfnamefont
  {F.~T.}\ \bibnamefont {Chong}},\ }\bibfield  {title} {\bibinfo {title}
  {Minimizing state preparations in variational quantum eigensolver by
  partitioning into commuting families},\ }\href
  {https://arxiv.org/abs/1907.13623} {\bibfield  {journal} {\bibinfo  {journal}
  {arXiv preprint arXiv:1907.13623}\ } (\bibinfo {year} {2019})}\BibitemShut
  {NoStop}%
\bibitem [{\citenamefont {Serra}\ and\ \citenamefont
  {Kais}(2000)}]{serra2000data}%
  \BibitemOpen
  \bibfield  {author} {\bibinfo {author} {\bibfnamefont {P.}~\bibnamefont
  {Serra}}\ and\ \bibinfo {author} {\bibfnamefont {S.}~\bibnamefont {Kais}},\
  }\bibfield  {title} {\bibinfo {title} {Data collapse for the
  {Schr{\"o}dinger} equation},\ }\href
  {https://doi.org/https://doi.org/10.1016/S0009-2614(00)00139-1} {\bibfield
  {journal} {\bibinfo  {journal} {Chem. Phys. Lett.}\ }\textbf {\bibinfo
  {volume} {319}},\ \bibinfo {pages} {273} (\bibinfo {year}
  {2000})}\BibitemShut {NoStop}%
\bibitem [{\citenamefont {Bhattacharjee}\ and\ \citenamefont
  {Seno}(2001)}]{bhattacharjee2001measure}%
  \BibitemOpen
  \bibfield  {author} {\bibinfo {author} {\bibfnamefont {S.~M.}\ \bibnamefont
  {Bhattacharjee}}\ and\ \bibinfo {author} {\bibfnamefont {F.}~\bibnamefont
  {Seno}},\ }\bibfield  {title} {\bibinfo {title} {A measure of data collapse
  for scaling},\ }\href {https://doi.org/10.1088/0305-4470/34/33/302}
  {\bibfield  {journal} {\bibinfo  {journal} {J. Phys. A: Math. Theor.}\
  }\textbf {\bibinfo {volume} {34}},\ \bibinfo {pages} {6375} (\bibinfo {year}
  {2001})}\BibitemShut {NoStop}%
\end{thebibliography}%

\clearpage
\pagebreak

\small

\section*{Methods}
\label{sec:methods}

\textbf{Quantum processors.} We utilized IBM quantum devices supporting mid-circuit measurements in our experiments. For experiments with projective measurements at $L \leq 5$, we used 7-qubit devices \textit{ibm\_lagos}, \textit{ibm\_perth}, \textit{ibmq\_jakarta}, and \textit{ibmq\_casablanca}. For $L > 5$, we used 27-qubit devices, \textit{ibm\_hanoi}, \textit{ibm\_cairo}, \textit{ibmq\_kolkata}, \textit{ibm\_auckland}, and a recent $127$-qubit device, \textit{ibm\_washington}. This latter set of devices supports sub-microsecond readout, with typical readout times $\sim \SI{750}{\nano\second}$, about twice as long as a CX gate and substantially faster than the $\sim \SI{5}{\micro\second}$ readout time on previous devices. For experiments with weak measurements, we used the 16-qubit device \textit{ibmq\_guadalupe} and the 27-qubit device \textit{ibmq\_montreal} \cite{jurcevic2021demonstration}. To maximize experiment throughput, we ran parallel experiments where possible. To avoid qubits with large gate errors and to limit cross-talk, we used at most two parallel circuits on sets of qubits separated by at least one idle qubit.

\textbf{Weak measurements.} We used null-type weak measurements \cite{gebhart2020topological, zilberberg2013null}, achieved by coupling the system qubit to an $\ket{0}$-ancilla through a unitary
\begin{equation}\begin{split}
    V(\eta) = e^{-i g \left(\mathbb{I} - \sigma^z\right) \otimes \sigma^y / 2}
    = \mqty[\mathbb{I} & 0 & 0 \\ 
        0 & \cos{g} & -\sin{g} \\
        0 & \sin{g} & \cos{g}],
\end{split}\end{equation}
where $\sin^2{g} = \eta$ and $\mathbb{I}$ is the $2 \times 2$ identity matrix. The effect of the coupling is
\begin{equation}\begin{split}
    &\left(a \ket{0} + b \ket{1}\right) \ket{0} \to \\
    & \qquad \left(a \ket{0} + b \sqrt{1 - \eta} \ket{1}\right) \ket{0}
        + b \sqrt{\eta} \ket{1} \ket{1}.
\end{split}\end{equation}

In the $\eta = 0$ weak limit, the coupling $V(\eta) = \mathbb{I}^{\otimes 2}$, and the system and ancilla qubits remain fully separable. In the $\eta = 1$ projective limit, $V(\eta)$ is maximally entangling, and measuring the ancilla provides complete information on the system qubit. Intermediate strengths $0 < \eta < 1$ smoothly interpolate between these two extremes. The Kraus operators describing the measurement are
\begin{equation}\begin{split}
    M^+(\eta) = \mqty[\dmat[0]{1, \sqrt{1-\eta}}], \qquad
    M^-(\eta) = \mqty[\dmat[0]{0, \sqrt{\eta}}].
\end{split}\end{equation}

\textbf{Circuit optimizations.} We applied several optimizations to the structure of our experiment circuits (\cref{fig:schematics-circuit}) to reduce circuit depth and complexity. Firstly, due to qubit connectivity constraints on hardware, our circuits are set on open chains of $L$ qubits instead of closed loops as in prior classical numerical studies \cite{szyniszewski2019entanglement, li2019measurement, zabalo2020critical}. Secondly, our randomized $2$-qubit gates each contain a single CX, in contrast with the 3 CXs required for general $2$-qubit operations, and on an individual level do not emulate Haar-uniform unitaries that have been used in prior studies \cite{szyniszewski2019entanglement, skinner2019measurement, bao2020theory}. We nonetheless remark that, taken collectively, the brickwork pattern of our $2$-qubit unitaries does approximate Haar-uniformity (see Supplementary Note~\inertlink{2}). The number of time steps in experiment circuits were chosen to be sufficient to reach $\gtrsim 95\%$ of steady-state saturation entanglement entropy (see Supplementary Note~\inertlink{1}). The specific structure of our circuits, for instance with each time step comprising two brickwork layers of randomized $2$-qubit gates and a measurement layer, was chosen to minimize the number of time steps required to reach steady-state.

An optimization to the decomposition of the coupling $V(\eta)$ in weak measurements was also used. An exact decomposition of $V(\eta)$ requires $2$ CX gates. However, as our ancillary qubit is always initialized to $\ket{0}$, it is necessary only for $V(\eta)$ to be exact in that sector. It is then possible to implement $V(\eta)$ with a single CX (see Supplementary Figure~\inertlink{1}). On circuits with multiple time steps, conditional resets \cite{corcoles2021exploiting} were used to re-initialize the ancilla to $\ket{0}$ after measurement, enabling re-use of the same set of ancillary qubits throughout the circuit. These resets operate by applying an $X$ gate conditional on a computational-basis measurement.

\textbf{Standard separable QST.} To recover an $n$-qubit density matrix $\rho$ through tomography, we employed projections onto all $2^n \times 3^n$ tensor products of $n$-qubit Pauli eigenstates, requiring $3^n$ quantum circuits each measuring a distinct Pauli string. This approach is referred to as standard separable quantum state tomography (SSQST), and has been shown to produce accurate tomography results compared to using non-overcomplete subsets of measurements \cite{nielsen2021gate, de2008choice}. To ensure physical $\rho$ estimates, we employed least-squares linear inversion with constraints of Hermiticity, unity trace, and positive semi-definiteness (PSD) \cite{Qiskit, smolin2012efficient}, instead of the computationally cheaper pseudoinverse.

\textbf{QST with simultaneous measurements.} At larger $n$, the number of tomography circuits and associated inversion costs can be greatly reduced by simultaneously measuring commuting Pauli strings \cite{hamamura2020efficient, yen2020measuring}. In particular, every set of $4^n - 1$ non-trivial Pauli strings on $n$ qubits can be partitioned into $2^n + 1$ groups, each containing $2^n - 1$ commuting strings. These groups are known as mutually unbiased bases (MUBs) \cite{durt2010mutually}. To determine MUBs for tomography, we exhaustively enumerated clique covers of the commutation graph of the Pauli strings, first excluding the $\{\mathbb{I}, \sigma^x\}^{\otimes n}$, $\{\mathbb{I}, \sigma^y\}^{\otimes n}$ and $\{\mathbb{I}, \sigma^z\}^{\otimes n}$ qubit-wise commuting families. As the number and size of cliques are fixed, the search procedure can be accelerated by recursion pruning, compared to the more general $\texttt{MIN-CLIQUE-COVER}$ problem. The clique covers generated enumerate the subclass of MUBs with maximal fully-separable basis sets \cite{romero2005structure}, yielding at least 3 tomography circuits that require no $2$-qubit gates. 

We used a stabilizer-based method to construct quantum circuits that implement MUB measurements \cite{gokhale2019minimizing}, selecting the lowest-depth circuits after a transpilation process that replaced all SWAP gates with classical swapping of readout data. From the relative outcome frequencies of the $n$ measurements on $n$ qubits of each circuit, covering $n$ Pauli strings, the expectation values of all $2^n - 1$ Pauli strings in the group can be recovered. Collecting $\expval{\sigma^{\vb*{\mu}}}_\rho$ for $\smash{\vb*{\mu} \in \{\mathbb{I}, x, y, z\}^{\otimes n}}$, we assemble
\begin{equation}\begin{split}
    A \vb*{\rho} = \vb{P}, \qquad
    A = \mqty[
        \mathbb{I}^{\otimes n} \\ 
        \vb*{\sigma}^{\vb*{\mu}_1} \\ 
        \vb*{\sigma}^{\vb*{\mu}_2} \\ 
        \vb*{\sigma}^{\vb*{\mu}_3} \\ \vdots], \qquad
    \vb{P} = \mqty[1 \\ \expval{\sigma^{\vb*{\mu}_1}}_\rho \\ \expval{\sigma^{\vb*{\mu}_2}}_\rho \\ \expval{\sigma^{\vb*{\mu}_3}}_\rho \\ \vdots],
\end{split}\end{equation}
for column-stacked vectorized density matrix $\vb*{\rho}$ and Pauli operators $\vb*{\sigma}^{\vb*{\mu}}$ row-wise flattened in $A$, such that the action of $\vb*{\sigma}^{\vb*{\mu}}$ on $\vb*{\rho}$ yields $\Tr[\rho \sigma^{\vb*{\mu}}]$, and perform a least-squares linear inversion with Hermiticity, unity trace, and PSD constraints to estimate $\vb*{\rho}$. We refer to this tomography strategy as MUBQST. Compared to SSQST, this approach yields significant savings in tomography costs with increasing $n$ (see Supplementary Table~\inertlink{1}). For example, at $n = 4$, an almost five-fold reductions in number of circuits is achieved. In our experiments, we used SSQST for $n \leq 2$ qubits, and MUBQST for $n > 2$ qubits. 

\textbf{Entropy mean and variance.} We computed entanglement entropy mean $\expval{S_\alpha}$ and variance $\var(S_\alpha)$ over all recorded trajectories on sampled instances of the hybrid random quantum circuits. Specifically, for a given $p$ and $\eta$, a set of experiment circuits $\{\mathcal{C}_i\}$ were generated and executed on hardware. On each circuit, each executed shot returns a mid-circuit measurement bitstring and a tomography measurement bitstring. The former records measurement outcomes over the time steps of the circuit and distinguishes the quantum trajectories realized on hardware. We categorized the tomography measurement outcomes by their mid-circuit bitstrings, enabling the recovery of the reduced density matrix and entanglement entropy of each trajectory. On circuit $\mathcal{C}_i$, we thus sampled a set of mid-circuit bitstrings $\mathcal{R}_i$, with associated entanglement entropy $S_\alpha[r]$ and relative frequency of occurrence $f[r]$ for each $r \in \mathcal{R}_i$. The entanglement entropy mean and variance are then calculated as
\begin{equation}\begin{split}
    \expval{S_\alpha} &= \frac{1}{\abs{\{\mathcal{C}_i\}}} 
        \sum_i \sum_{r \in R_i} f[r] S_\alpha[r], \\
    \var(S_\alpha) &= \frac{1}{\abs{\{\mathcal{C}_i\}}} 
        \sum_i \sum_{r \in R_i} f[r] \left(S_\alpha[r] - \expval{S_\alpha}\right)^2.
\end{split}\end{equation}

Due to their lower hardware resource costs, for $L = 3$ with weak measurements and $L = 4, 5$ with projective measurements, we sampled $\gtrsim 300$ and $\gtrsim 100$ random circuits for each $(p, \eta)$ datapoint, respectively; for all other $L$ we sampled $\geq 50$ random circuits per datapoint. There are $\sim 10$ MUBs for each sampled random circuit (see Supplementary Table~\inertlink{1}) and we executed $\gtrsim 1000 \times 2^{m}$ shots per MUB for tomography, where $m$ is the number of mid-circuit measurements in the circuit. Within an experiment, trajectories appearing with $\leq 128$ total shots are discarded as they present insufficient data for reliable tomography. The number of mid-circuit measurements $m$ in circuits range from $\sim 4$ to $\sim 14$ depending on $L$ (see Supplementary Table~\inertlink{2}).

\textbf{Subsystem size.} To probe entanglement crossovers with projective and weak measurements (\cref{fig:results-proj-trans-L4-mean,fig:results-proj-trans-L5-mean,fig:results-proj-trans-L4-var,fig:results-proj-trans-L5-var,fig:figure-results-weak}) and criticality of the transition (\cref{fig:results-collapse}), we used subsystem size $\abs{A} = \floor{L/2}$. In probing the scaling of entanglement entropy with system size (\cref{fig:results-proj-scaling}), however, we used quarter subsystem size to reduce hardware resource costs in our experiments. To produce a smooth curve, we interpolated the value of $\expval{S_2}$ between $\abs{A} = \floor{L/4}$ and $\abs{A} = \ceil{L/4}$. The alternate choice of only taking $\abs{A} = \floor{L/4}$ introduces step-like patterns in $\expval{S_2}$ versus $L$, but the key observation of extensive and sub-extensive scaling of entanglement entropy in volume- and area-law phases is unchanged.

\textbf{Readout error mitigation (RO).} Measurement bit-flip error rates were acquired through calibration circuits run alongside experiments, and we performed least-squares linear inversion on raw measurement counts to approximately correct these errors \cite{jurcevic2021demonstration, kandala2017hardware}. In cases where only a subset of the $n$ qubits are measured, occurring in $p < 1$ time steps and during QST on subsystems, the reduced calibration matrix is obtained by partial summation over qubits that are not measured. We used complete readout mitigation for $n \leq 5$ qubits, and tensored readout mitigation \cite{Qiskit} for $n > 5$. The latter procedure calibrates qubit sub-registers in parallel, reducing the number of calibration circuits at an expense of neglecting correlations in readout error between qubits in different sub-registers.

\textbf{Residual entropy correction (RE).} On the random quantum circuits in our experiments, coherent errors from the quantum gates are effectively averaged out into incoherent noise. Then, in the presence of hardware noise, the measured entanglement entropy is $S_\alpha = S_\alpha^* + \delta S_\alpha$, with $S_\alpha^*$ being the true value and $\delta S_\alpha$ arising from the combined decoherence. By taking $p = 1$ as the zero reference, we may calibrate the anomalous contribution $\delta S_\alpha$ and remove it from experiment data. To do so, we approximated that $\delta S_\alpha(p, \eta)$ scales linearly with the error of quantum circuits $\mathcal{C}$ at measurement rate $p$ and strength $\eta$, $\expval{\mathcal{E}[\mathcal{C}_{p,\eta}]}$, as calculated by composing reported gate-level error rates. The average is taken over all sampled circuits $\mathcal{C}$ for the experiment. The circuit error was estimated as
\begin{equation}\begin{split}
    \mathcal{E}[\mathcal{C}] \approx \max_j
        \left(\varepsilon^{\text{1q}} N^{\text{1q}}_j [\mathcal{C}]
        + \varepsilon^{\text{2q}} N^{\text{2q}}_j [\mathcal{C}]
        + \varepsilon^{\text{ro}} N^{\text{ro}}_j [\mathcal{C}]\right),
\end{split}\end{equation}
where $\smash{N^{\text{1q}}_j [\mathcal{C}], N^{\text{2q}}_j [\mathcal{C}], N^{\text{ro}}_j [\mathcal{C}]}$ are the number of 1-qubit, 2-qubit CX, and measurement gates respectively in circuit $\mathcal{C}$ involving qubit $j$, and $\varepsilon^{\text{1q}}, \varepsilon^{\text{2q}}, \varepsilon^{\text{ro}}$ are 1-qubit, 2-qubit CX, and measurement error rates respectively, as reported in hardware calibration data. This linear scaling of $\delta S_\alpha(p, \eta)$ with circuit error is supported by characterization data (see Supplementary Note~\inertlink{3}). Note $\mathcal{C}$ includes tomography gates and measurements, appended to the end of the experiment circuit. 

The final quantum state should always be fully disentangled at $p = \eta = 1$, since all qubits are measured projectively at every time step. Any residual entanglement entropy detected must arise entirely from decoherent errors, that is, $S_\alpha(p = \eta = 1) = \delta S_\alpha(p = \eta = 1) \geq 0$. Thus, the anomalous entropy $\delta S_{\alpha}$ is related to the $\delta S_\alpha(p = \eta = 1) \geq 0$ reference as
\begin{equation}\begin{split}
    \delta S_\alpha(p, \eta) &= \frac{\expval{\mathcal{E}[\mathcal{C}_{p,\eta}]}}{\expval{\mathcal{E}[\mathcal{C}_{p = \eta = 1}]}} S_\alpha(p = \eta = 1).
\end{split}\end{equation}

Subtracting this correction from $S_\alpha(p, \eta)$ removes the anomalous entropy and provides a better estimate of the true entanglement entropy. Note there are no free parameters; in all experiments where RE is applied, data for $S_\alpha(p = \eta = 1)$ were either part of the presented dataset or collected separately. The $7$-qubit processors and the $\geq 27$-qubit processors utilized in our experiments differ in their error characteristics. We took $\varepsilon^{\text{1q}} \approx 3 \times 10^{-4}$ and $\varepsilon^{\text{2q}} \approx 4 \times 10^{-3}$ for both families, but $\varepsilon^{\text{ro}} \approx 5 \times 10^{-3}$ for the former and $\varepsilon^{\text{ro}} \approx 8 \times 10^{-2}$ for the latter. These error rates reflect values seen on the qubit chains picked by our selection algorithm and may not be representative of the average error on the processors. We checked additionally the robustness of our main results to data post-processing in Supplementary Notes~\inertlink{4} and~\inertlink{5}, in particular using the trivial RE scheme $\delta S_\alpha(p, \eta) = S_\alpha(p = \eta = 1)$ with no scaling corrections to entropic contribution from decoherence, in place of the linear scheme. 

\textbf{Qubit selection.} The quantum processors we utilized have more qubits than necessary for our experiments, thus presenting choices for the selection of qubits on which circuits are executed. We picked the set of qubits that minimize estimated circuit error. Specifically, given the set $\mathcal{X}$ of qubit selections that satisfy connectivity constraints, we estimated the mean error
\begin{equation}\begin{split}
    \mathcal{E}_{x \in \mathcal{X}} [\mathcal{C}] = \sum_{j \in x} \left(\varepsilon^{\text{1q}}_j N^{\text{1q}}_j [\mathcal{C}] 
        + \varepsilon^{\text{2q}}_j N^{\text{2q}}_j [\mathcal{C}]
        + \varepsilon^{\text{ro}}_j N^{\text{ro}}_j [\mathcal{C}] \right),
\end{split}\end{equation}
where $\varepsilon^{\text{1q}}_j, \varepsilon^{\text{2q}}_j, \varepsilon^{\text{ro}}_j$ are 1-qubit, 2-qubit CX, and measurement error rates respectively on qubit $j$, as reported in hardware calibration data. We selected the set of qubits $x$ minimizing $\expval{\mathcal{E}_x[\mathcal{C}]}$, where the average is taken over the circuits to be executed, including their tomography components. We performed the minimization by an exhaustive search over the qubit connectivity graph of the processors. For weak measurements, qubits are selected to place an ancillary qubit adjacent to each system qubit (see Supplementary Figure~\inertlink{1}).

\textbf{Collapse of hardware data.} The data collapse procedure proceeds by way of numerical minimization of a measure of the scatter of data points, yielding the best-fit critical exponents $\gamma$, $\nu$. The critical measurement rate $p^*$ can either be determined separately or simultaneously with $\gamma$, $\nu$ in the collapse. We took $p^* \approx 0.25$ from hardware $\var(S_1)$ peak at $L = 5$ with projective measurements (\cref{fig:results-proj-trans-L5-var}) as well as $p^* \approx 0.22$ from an extrapolation scheme into the thermodynamic limit (see Supplementary Note~\inertlink{5}) to check robustness. A complication is that the scaling function $F$, which sets the curve onto which the data collapses, is unknown, and the procedure must depend only on the experiment dataset \cite{skinner2019measurement, serra2000data, bhattacharjee2001measure}. To avoid preferential treatment of any portion of the data, we define the measure of scatter symmetrically over each data subset \cite{bhattacharjee2001measure}. We denote by $\mathcal{L}$ the set of system sizes in experiments, and $\mathcal{P}_L$ the set of sampled measurement rates $p$ for each $L$. From experiment data $\smash{\left(p, \expval{S_\alpha(p)}\right)}$ at $L \in \mathcal{L}$ and $p \in \mathcal{P}_L$, we compute the pairs $\smash{\left(q_L(p), W_L(p)\right)}$ rescaled as
\begin{equation}\begin{split}
    q_L(p) &= (p - p^*) L^{1 / \nu}, \\
    W_L(p) &= \left(\expval{S_\alpha(p)} - \expval{S_\alpha(p^*)}\right) L^{-\gamma / \nu},
\end{split}\end{equation}
which follows from the finite-size scaling form in the main text. With the rescaled data at each $L$, we constructed a smoothed interpolating function $f_L(q)$, such that $f_L(q)$ gives the rescaled entanglement entropy $W_L$ at each $q_L$. We denote the set of $q_L$ as $\mathcal{Q}_L$, and $\smash{q_L^- = \min{\mathcal{Q}_L}}$ and $\smash{q_L^+ = \max{\mathcal{Q}_L}}$. Adapting the measure of goodness of fit from \cite{bhattacharjee2001measure}, we define the loss function
\begin{equation}\begin{split}
    R(\gamma, \nu) = 
        \sum_{L \in \mathcal{L}} \sum_{\substack{L' \in \mathcal{L} \\ L' \neq L}} \,
        L^{2 \gamma / \nu} 
        \sum_{\substack{q \in \mathcal{Q}_{L'} \\ q_L^- \leq q \leq q_L^+}} 
        \left[f_L(q) - f_{L'}(q)\right]^2.
\end{split}\end{equation}

The loss $R(\gamma, \nu)$ is a sum of squared residuals, measuring the scatter of rescaled data against the curve presented by the subset of data at each $L$. In assessing the scatter of the rescaled data, only data within a $q$-interval overlapping with the considered curve is considered \cite{skinner2019measurement, bhattacharjee2001measure}, that is, $\smash{q \in [q_L^-, q_L^+]}$. This procedure avoids extrapolation of $f_L(q)$. The prefactor of $\smash{L^{2 \gamma / \nu}}$ is for normalization after rescaling. The best-fit critical exponents are then $\smash{(\gamma_0, \nu_0) = \argmin_{\gamma, \nu} R(\gamma, \nu)}$. Estimates of fitting errors are set by the width of the minimum; for simplicity we consider the $\gamma$ and $\nu$ parameters at the quadratic level \cite{bhattacharjee2001measure},
\begin{equation}\begin{split}
    \delta \gamma^\pm &= \epsilon \gamma_0 \left[ 2 \ln 
        \frac{R(\gamma_0 \pm \epsilon \gamma_0, \nu_0)}{R(\gamma_0, \nu_0)} \right]^{-1/2}, \\
    \delta \nu^\pm &= \epsilon \nu_0 \left[ 2 \ln 
        \frac{R(\gamma_0, \nu_0 \pm \epsilon \nu_0)}{R(\gamma_0, \nu_0)} \right]^{-1/2}.
\end{split}\end{equation}

In our analysis, we considered the four largest system sizes $\mathcal{L} = \{5, 6, 7, 8\}$ for which experiment data is available. These sizes were chosen to limit finite-size effects, which at small $L$ causes deviations from the scaling form. To ensure the correct identification of the global minimum, a grid search followed by gradient descent was used. We report conservative estimates with lumped error $\gamma_0 \pm \max{(\delta \gamma^+, \delta \gamma^-)}$ and $\nu_0 \pm \max{(\delta \nu^+, \delta \nu^-)}$, at the $\epsilon = 1\%$ level following Ref.~\cite{bhattacharjee2001measure}.

\clearpage



\section*{Supplementary material}

# Experimental Realization of a Measurement-Induced Entanglement Phase Transition on a Superconducting Quantum Processor
## (Supplementary Information)

Jin Ming Koh,[1] Shi-Ning Sun,[2] Mario Motta,[3] and Austin J. Minnich[2, *]

[1] *Division of Physics, Mathematics and Astronomy,
California Institute of Technology, Pasadena, California 91125, USA*
[2] *Division of Engineering and Applied Science, California Institute of Technology, Pasadena, CA 91125, USA*
[3] *IBM Quantum, IBM Research Almaden, San Jose, CA 95120, USA*

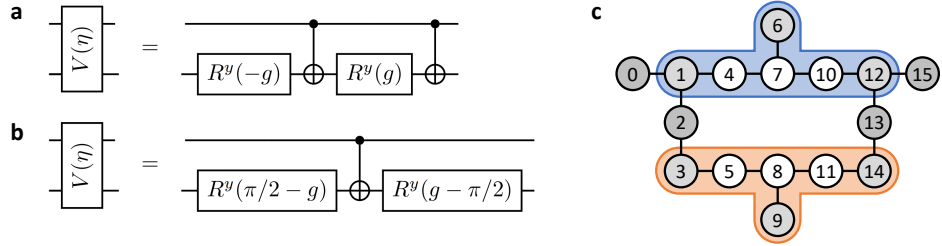

Supplementary Figure 1. Implementation of $V(\eta)$ for weak measurements and selection of qubits on hardware. (a) Exact decomposition of $V(\eta)$ for arbitrary input, and (b) decomposition of $V(\eta)$ exact only in the ancilla-$|0\rangle$ sector, using only a single CX gate. (c) Illustration of $3+3$ qubit placements (dark blue and light orange highlighted groups) for $L=3$ quantum circuits with weak measurements, on 16-qubit quantum device *ibmq_guadalupe*. Weak measurement on a system qubit requires an adjacent ancilla, thus constraining possible ideal qubit selections. In the highlighted groups, qubits in the middle shaded white are the system qubits, and surrounding qubits shaded light gray are the ancillae.

|  | Number of Tomography Circuits | |
|---|---|---|
| Subsystem Size $|A|$ | SSQST | MUBQST |
| 1 | 3 | - |
| 2 | 9 | 5 |
| 3 | 27 | 9 |
| 4 | 81 | 17 |
| 5 | 343 | 33 |

Supplementary Table 1. Comparison of number of tomography measurement circuits needed for SSQST and MUBQST for a given trajectory defined by mid-circuit measurement outcomes. To obtain the reduced density matrix $\rho_A$ over the subsystem of a single hybrid random circuit on a given trajectory, the number of distinct tomography circuits listed above needs to be executed to measure a complete set of Pauli expectation values. MUBQST exploits simultaneous measurements on commuting groups of Pauli strings and requires fewer tomography circuits than SSQST.

* aminnich@caltech.edu

**Supplementary Note 1: Trajectories of entanglement entropy versus time**

We provide an illustration of the dynamical behaviour of entanglement entropy on our hybrid random quantum circuits. As discussed, at small measurements rates $p$ or strengths $\eta$, a volume-law phase manifests, and a ballistic growth in the entanglement entropy of a subsystem occurs before saturation in time [1, 2]. On the other hand, at large $p$ or $\eta$, an area-law phase manifests, and no ballistic growth occurs. On a 1D system, as investigated in this work, the entanglement entropy thus remains near zero throughout. We illustrate these dynamical features in Supplementary Figure 2, which shows individual realizations of von Neumann entanglement entropy $S_1$ against time steps of the circuit, computed by noiseless classical emulation. In our experiments, we select a number of time steps sufficient to reach $\gtrsim 95\%$ of steady-state entanglement entropy. We present the average number of time steps used at each $L$ in Supplementary Table 2.

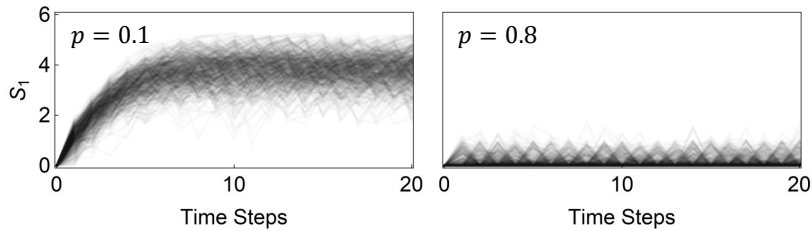

Supplementary Figure 2. Trajectories of entanglement entropy $S_1$ against time, computed numerically at system size $L = 12$, subsystem size $|A| = L/2$. Each translucent black line tracks an individual realization; 500 realizations are plotted. Similar dynamical behaviour occurs for higher-order Rényi entanglement entropies.

| System Size $L$ | Avg. # Time Steps | Max. # Mid-Circuit Measurements |
|---|---|---|
| 2 | 2 | 4 |
| 3 | 2 | 6 |
| 4 | 2 | 8 |
| 5 | 3 | 10 |
| 6 | 3 | 12 |
| 7 | 4 | 12 |
| 8 | 4 | 14 |
| 10 | 4 | 14 |
| 12 | 5 | 14 |
| 14 | 5 | 14 |

Supplementary Table 2. Average number of time steps and maximum number of mid-circuit measurements in sampled hybrid random quantum circuits across $p$ at various system sizes $L$. Note $L = 10, 12, 14$ were sampled only at $p = 0.1, 0.8, 1$ to probe the scaling of entanglement entropy with system size, and an average $\sim 4$–$5$ time steps were sufficient to reach $\gtrsim 95\%$ of steady-state entanglement entropy at these $p$.

**Supplementary Note 2: Comparison of brickwork 2-qubit gates against Haar-uniform unitaries**

As a number of prior numerical works utilize Haar-uniform entangling unitaries in their circuit models [1–3], it is of interest to compare the unitaries realized by our randomized 2-qubit gates, which contain a single CX each to minimize circuit depth and are tessellated in a brickwork pattern, against Haar-uniform implementations. As a diagnostic, we examine the eigenphase and normalized eigenphase spacing distributions of $n$-qubit unitaries realized by our 2-qubit gates in brickwork pattern, following Ref. [4]. Specifically, given eigenvalues $\{e^{i\theta_1}, e^{i\theta_2}, \ldots, e^{i\theta_N}\}$ of a sampled unitary, $N = 2^n$, we extract the phases $-\pi \leq \theta_1 \leq \theta_2 \leq \ldots \leq \theta_N < \pi$, and compute their normalized spacings

$$\left\{ s_j = \frac{N}{2\pi}(\theta_{j+1} - \theta_j) \mid j = 1, \ldots, N-1 \right\}. \tag{1}$$

We present in Supplementary Figure 3a numerically sampled histograms comparing the relative frequencies of eigenphases and normalized eigenphase spacings for $n$-qubit unitaries generated by our 2-qubit gates in brickwork pattern, and those from $n$-qubit Haar-uniform unitaries, at $n = 4, 8$. The distribution of eigenphases for Haar-uniform unitaries is expected to be uniform over $[-\pi, \pi)$, as indeed observed. The quantitative similarity between the eigenphase and eigenphase spacing distributions suggest that our 2-qubit gate implementations does approximate Haar-uniformity. For a further comparison, we replace all 2-qubit gates in our circuits with Haar-uniform 2-qubit unitaries and repeat our ideal noiseless numerical emulations. We present the mean and variance of von Neumann entanglement entropy $S_1$ against $p$ in Supplementary Figure 3b, comparing curves using our 2-qubit unitaries and Haar-uniform unitaries. Slight differences are observed, but the overall similarity between the two sets of results again indicate that our 2-qubit gates in brickwork pattern adequately approximate Haar-uniformity for the purposes of this study.

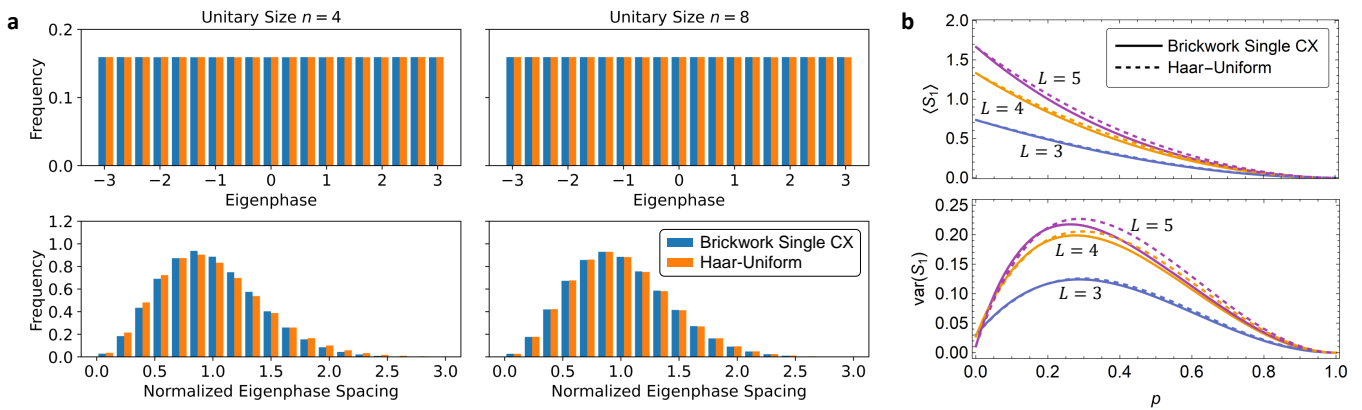

Supplementary Figure 3. Comparison of brickwork pattern of single-CX 2-qubit gates with Haar-uniform unitaries. (a) Numerically sampled histograms showing relative frequencies of eigenphases and normalized eigenphase spacings, for $n = 4$ and $n = 8$ qubit unitaries generated by our 2-qubit gates in brickwork pattern, and unitaries sampled uniformly from the Haar measure. (b) Noiseless numerics of the mean and variance of entanglement entropy $S_1$ versus $p$ on our experiment circuits, and on circuits with all 2-qubit gates replaced by Haar-uniform unitaries, at system sizes $L = 3, 4, 5$.

**Supplementary Note 3: Residual entropy characterization**

On the random quantum circuits used in our experiments, coherent errors from the quantum gates are expected to average into an effective decoherence-like effect. Our residual entropy (RE) correction scheme assumes a linear scaling $\delta S_\alpha(p, \eta) \propto \langle \mathcal{E}[\mathcal{C}_{p,\eta}] \rangle$, that is, the entropic contributions from decoherence are proportional to the estimated circuit error. For sufficiently small per-gate errors, error addition trivially motivates this relation. Here we present characterization data, obtained on the same quantum devices as used in our experiments, that directly support the assumed linearity.

The estimation of $\langle \mathcal{E}[\mathcal{C}] \rangle$ for an experiment circuit $\mathcal{C}$ includes also the tomography gates and measurements, appended onto the time-stepping simulation circuit. We write $\langle \mathcal{E}[\mathcal{C}_{p,\eta,T}] \rangle = \langle \mathcal{E}[\mathcal{C}_{p,\eta,T}] \rangle_{\text{sim}} + \langle \mathcal{E}[\mathcal{C}] \rangle_{\text{tomo}}$, where the contribution from tomography $\langle \mathcal{E}[\mathcal{C}] \rangle_{\text{tomo}}$ is dependent only on the tomography specifics, namely, whether SSQST or MUBQST is used, and subsystem size $|A|$ on which tomography is performed. In particular, $\langle \mathcal{E}[\mathcal{C}] \rangle_{\text{tomo}}$ is independent of measurement rate $p$ and strength $\eta$, and the number of simulation time steps $T$. As described in Methods, we calculate circuit errors $\mathcal{E}[\mathcal{C}]$ by summing gate errors over the depth of the circuit. Each time-step is equivalent in depth, so $\langle \mathcal{E}[\mathcal{C}_{p,\eta,T}] \rangle_{\text{sim}} \propto T$. Assuming $\delta S_\alpha \propto \langle \mathcal{E}[\mathcal{C}] \rangle$, we thus expect

$$\langle \delta S_\alpha(p, \eta, T) \rangle - \langle \delta S_\alpha(p, \eta, T = 0) \rangle \propto T, \tag{1}$$

where $T = 0$ means the simulation circuit contains only the tomography components. At $p = \eta = 1$, the final quantum state should always be fully disentangled. Then any residual entanglement entropy must arise from decoherence, $S_\alpha(p = \eta = 1) = \delta S_\alpha(p = \eta = 1)$. Thus, at $p = \eta = 1$, we expect

$$\langle S_\alpha(T) \rangle - \langle S_\alpha(T = 0) \rangle \propto T. \tag{2}$$

We present in Supplementary Figure 4 measurements of $\langle S_\alpha(T) \rangle - \langle S_\alpha(T = 0) \rangle$ against $T$, for differing system sizes $L$, and subsystem sizes $|A|$ on which tomography is performed. Quantum state tomography for $|A| \leq 2$ was performed using SSQST, and for $|A| > 2$ was performed using MUBQST. The figure shows that after readout error mitigation (RO), all characterization experiments exhibit quantitative linear trends of residual entanglement entropy against $T$. This trend is observed for both Rényi orders $\alpha = 1$ and $\alpha = 2$, computed from the same dataset. This set of characterization data, acquired separately from our experiments, thus directly supports the $\delta S_\alpha \propto \langle \mathcal{E}[\mathcal{C}] \rangle$ linear relation assumed. Effectively, the RE correction scheme is analogous to a calibration of the quantum device such that entropic contributions from averaged errors are zeroed out based on $p = \eta = 1$ references.

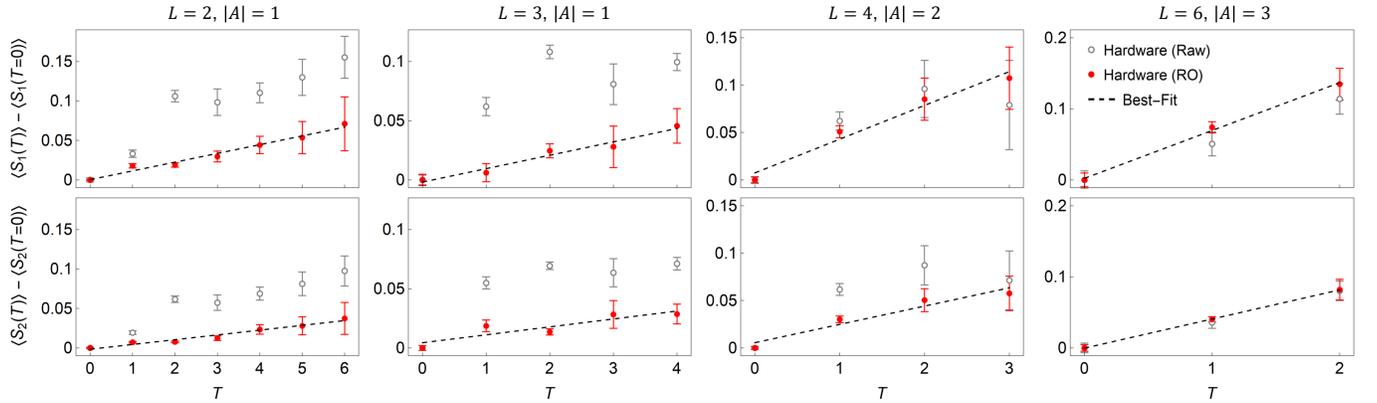

Supplementary Figure 4. Measured scaling of residual entanglement entropy with circuit depth. Entanglement entropies at Rényi orders $\alpha = 1$ and $\alpha = 2$ are observed to be proportional to circuit depth, supporting the linearity assumption in our RE correction scheme. Data for $L \leq 4$ were obtained on 7-qubit quantum devices {*ibm_lagos, ibmq_jakarta, ibmq_casablanca, ibm_perth*}, and $L > 4$ were obtained on 27-qubit devices {*ibm_hanoi, ibmq_cairo, ibmq_kolkata*}. Error bars reflect 90% confidence intervals estimated from statistical bootstrapping.

**Supplementary Note 4: Effect of relaxed RE correction on entanglement entropy under projective measurements and scaling of entanglement entropy with system size**

Our main experimental results exhibit quantitative agreement with theoretical expectations when RO and RE mitigations are applied, as shown in Figures 2–4 of the main text. To assess the dependence of our conclusions on this procedure, we consider other data post-processing procedures. Measured entanglement entropies without RO and RE, and with RO only, had been reported in Figures 2 and 3 of the main text. Due to the non-negligible readout error rates on hardware ($\geq 1\%$ typical) and the large number of gates on the random circuits whose averaged errors present non-negligible entropic contributions, removing RO and RE degrades the agreement with theoretical expectations. Nonetheless, the important qualitative features of the data are preserved.

Here, we present the same experiment data as in the main text but processed with a relaxed form of RE correction that uses the trivial scheme $\delta S_\alpha(p, \eta) = S_\alpha(p = \eta = 1)$, instead of the linear $\delta S_\alpha(p, \eta) \propto \langle \mathcal{E}[\mathcal{C}_{p,\eta}] \rangle$ of the full scheme. That is, the relaxed scheme ignores the scaling of residual entropy with depth-dependent circuit error, and subtracts a uniform entropy $S_\alpha(p = \eta = 1)$ from all data at the same $L$ and $|A|$. We first present the analogue of Figure 2a of the main text, which reports the mean and variance of $S_1$ against $p$ under projective measurements at $L = 4, 5$, in Supplementary Figure 5a. The data with relaxed RE is consistent to within 4% of that with full RE, and a quantitative agreement with theory is maintained. In Supplementary Figure 5b, corresponding to Figure 2b of the main text, the scaling of $\langle S_2 \rangle$ against system size $L$ also exhibits a quantitative match with theory using the relaxed RE scheme. Importantly, the volume- and area-law scaling of entanglement entropy is evident.

The insensitivity of results in Supplementary Figure 5a to the RE scheme is largely because at smaller system sizes $L$, variations in circuit depth, and hence circuit error, with $p$ are small. Likewise, that in Supplementary Figure 5b is because circuit depths at $p = 0.1, 0.8, 1$ are similar. Generally, the full RE scheme becomes important as $L$ increases, at intermediates values of $p$. Nonetheless, we found that data collapse onto a single curve is still qualitatively observed with the relaxed RE scheme (see Supplementary Note 5).

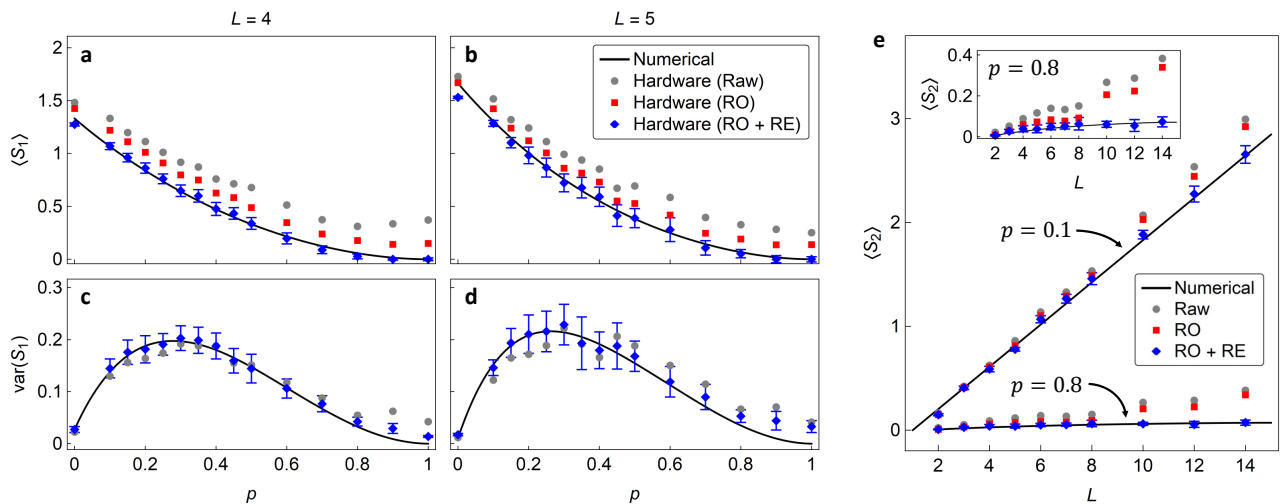

Supplementary Figure 5. Entanglement crossover and system size scaling under projective measurements, using relaxed RE correction. Average von Neumann entanglement entropy $\langle S_1 \rangle$ versus measurement rate $p$ at system sizes **(a)** $L = 4$ and **(b)** $L = 5$, and $S_1$ entanglement entropy variance versus $p$ at **(c)** $L = 4$ and **(d)** $L = 5$, obtained on quantum hardware. **(e)** Scaling of second-order Rényi entanglement entropy against system size, up to $L = 14$ qubits, obtained on hardware. Quantum devices {*ibm_lagos, ibm_perth, ibmq_jakarta, ibmq_casablanca*} were used for $L \leq 5$, and {*ibm_hanoi, ibm_cairo, ibm_kolkata, ibm_auckland, ibm_washington*} with sub-microsecond readout were used for $L > 5$. Error bars reflect 90% confidence intervals estimated from statistical bootstrapping.

**Supplementary Note 5: Data collapse results using $p^*$ extrapolated into the thermodynamic limit and relaxed RE correction**

As described in the main text, instead of using $p^* \approx 0.25$ estimated from experimentally-measured $S_1$ variance peaks at $L = 4, 5$, which is indicative of the crossover point between volume- and area-law phases (Figure 2a in the main text), an extrapolation scheme into the thermodynamic limit can be used. The $S_1$ variance peaks in ideal numerical data up to $L = 16$, at half subsystem size, are identified, and a linear regression is performed; then $p^*$ in the thermodynamic limit ($1/L^2 \to 0$) can be obtained. This approach yields $p^* \approx 0.22$, and performing data collapse in an identical fashion as in the main text (Supplementary Figure 6a, gives best-fit critical exponents $\gamma \approx 2.3 \pm 0.4$ and $\nu \approx 2.3 \pm 0.3$. These are consistent with the exponents obtained using experimentally-determined $p^* \approx 0.25$ in the main text, to within uncertainties.

We also investigate data collapse using the relaxed RE scheme $\delta S_\alpha(p, \eta) = S_\alpha(p = \eta = 1)$, instead of the full scheme. The collapsed entanglement entropies are shown in Supplementary Figure 6b, using $p^* \approx 0.25$. As in the main text, the solid black line shows ideal numerical results at $L = 16$ under the same rescaling, for comparison. The best-fit critical exponents obtained are $\gamma \approx 5.0 \pm 1.0$ and $\nu \approx 3.3 \pm 0.5$, which deviate from those obtained using the full RE scheme in main text. However, the important qualitative observation that the hardware data collapses approximately onto a single curve remains evident. The conclusion that a critical measurement-induced quantum phase transition had been realized is thus unchanged. Together, the findings in Supplementary Figures 5 and 6 demonstrate that our observation of a quantum phase transition is robust to different post-processing procedures.

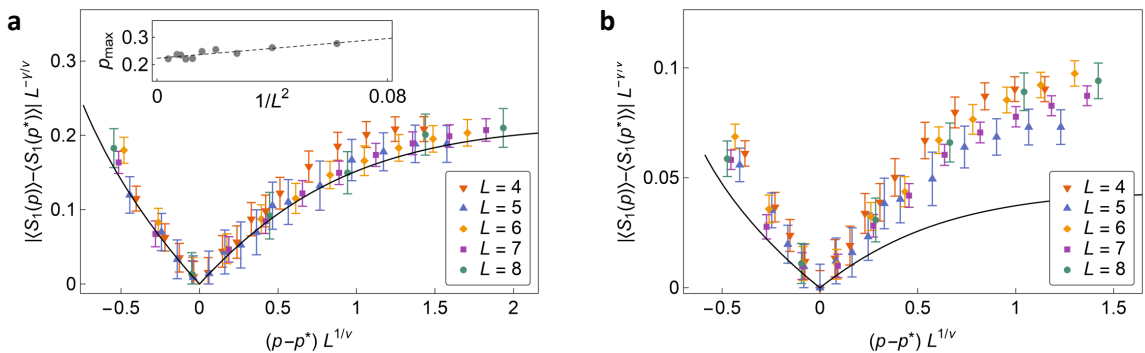

Supplementary Figure 6. Data collapse results using alternate post-processing procedures. (a) Collapsed average Von Neumann entanglement entropies using $p^* \approx 0.22$, determined through an extrapolation of numerical entropy variance peaks towards the thermodynamic limit. (b) Collapsed average Von Neumann entanglement entropies using relaxed RE correction, which ignores variations in residual entropy with depth-dependent circuit error. Solid line shows ideal numerical results at $L = 16$, rescaled using best-fit $(\gamma, \nu)$ determined from hardware data, for comparison. Quantum devices {*ibm_lagos, ibm_perth, ibmq_jakarta, ibmq_casablanca*} were used for $L \leq 5$, and {*ibm_hanoi, ibm_cairo, ibmq_kolkata*} were used for $L > 5$. Error bars reflect 90% confidence intervals estimated from statistical bootstrapping.

[1] B. Skinner, J. Ruhman, and A. Nahum, Measurement-induced phase transitions in the dynamics of entanglement, [Phys. Rev. X **9**, 031009 (2019)](#).
[2] M. Szyniszewski, A. Romito, and H. Schomerus, Entanglement transition from variable-strength weak measurements, [Phys. Rev. B **100**, 064204 (2019)](#).
[3] Y. Bao, S. Choi, and E. Altman, Theory of the phase transition in random unitary circuits with measurements, [Phys. Rev. B **101**, 104301 (2020)](#).
[4] F. Mezzadri, How to generate random matrices from the classical compact groups, [arXiv preprint math-ph/0609050 (2006)](#).



\end{document}